\documentclass[12pt,preprint]{aastex}

\newcommand{\tsq}{$t^{2}$}
\newcommand{\cmsq}{cm$^{-2}$}
\newcommand{\s}{s$^{-1}$}
\newcommand{\tsqa}{$t^{2}_A$}
\newcommand{\Te}{\mbox{$T_{e}$}}
\newcommand{\Tc}{\mbox{$T_{c}$}}
\newcommand{\Ha}{H$\alpha$}
\newcommand{\Hb}{H$\beta$}
\newcommand{\Hg}{H$\gamma$}
\newcommand{\HII}{\mbox{H~{\sc ii}}}
\newcommand{\HI}{\mbox{H~{\sc i}}}
\newcommand{\OIII}{\mbox{[O~{\sc iii}]}}
\newcommand{\OII}{\mbox{[O~{\sc ii}]}}
\newcommand{\OI}{\mbox{[O~{\sc i}]}}
\newcommand{\NII}{\mbox{[N~{\sc ii}]}}
\newcommand{\etal}{et al.}
\newcommand{\thC}{$\theta^1$~Ori~C}
\newcommand{\thtwo}{$\theta^2$~Ori~A}
\newcommand{\cmq}{ cm$^{-3}$}
\newcommand{\Ncrit}{$N_{crit}$}
\newcommand{\tfour}{$T_{4}$}
\newcommand{\Ne}{$N_{e}$}
\newcommand{\Fe}{\mbox{[Fe~{\sc iii}]}}
\newcommand{\He}{\mbox{He~{\sc i}}}
\newcommand{\Whb}{$W_{\rm H\beta}$}
\newcommand{\hplus}{\mbox{H$\rm ^{+}$}}
\newcommand{\heneutral}{\mbox{He$\rm ^{o}$}}
\newcommand{\heplus}{\mbox{He$\rm ^{+}$}}

\newcommand{\SII}{\mbox{[S~{\sc ii}]}}
\newcommand{\SIII}{\mbox{[S~{\sc iii}]}}
\newcommand{\kms}{km~s{$^{-1}$}}
\newcommand{\LyC}{\mbox{LyC}}
\newcommand{\htwo}{H$_{2}$}
\newcommand{\subsun}{M$_{\hbox{$\odot$}}$}
\newcommand{\yr}{yr{$^{-1}$}}

\shorttitle{Fine Scale Temperature Fluctuations in the the Orion Nebula}
\shortauthors{O'Dell, Peimbert \&\ Peimbert}

\begin{document}

\title{Fine Scale Temperature Fluctuations in the the Orion Nebula and the
\tsq\ Problem
\footnote{
Based in part on observations with the NASA/ESA Hubble Space Telescope,
obtained at the Space Telescope Science Institute, which is operated by
the Association of Universities for Research in Astronomy, Inc., under
NASA Contract No. NAS 5-26555.}}

\author{C. R. O'Dell}
\affil{Department of Physics and Astronomy, Vanderbilt University,
Box 1807-B, Nashville, TN 37235}
\email{cr.odell@vanderbilt.edu}
\author{Manuel Peimbert}
\affil{Instituto de Astronom\'{\i}a, Apartado Postal 70-264, UNAM, Mexico City, DF 004510 M\'exico}
\author{Antonio Peimbert}
\affil{Instituto de Astronom\'{\i}a, Apartado Postal 70-264, UNAM, Mexico City, DF 004510 M\'exico}

\begin{abstract}

We present a high spatial resolution map of the columnar electron temperature
(\Tc) of a region to the south west of the Trapezium in the Orion
Nebula. This map was derived from Hubble Space Telescope images that
isolated the primary lines of \HI\ for determination of the local
extinction and of the \OIII\ lines for determination of \Tc.  Although
there is no statistically significant variation of \Tc\ with distance
from the dominant ionizing star \thC, we find small scale variations
in the plane of the sky down to a few arcseconds that are compatible
with the variations inferred from comparing the value of \Te\ derived
from forbidden and recombination lines, commonly known as the \tsq\
problem.  We present other evidence for fine scale variations in
conditions in the nebula, these being variations in the surface
brightness of the the nebula, fluctuations in radial velocities, and
ionization changes. From our \Tc\ map and other considerations we
estimate that \tsq\ $= 0.028 \pm 0.006$ for the Orion nebula. Shadowed
regions behind clumps close to the ionization front can make a significant
contribution to the observed temperature fluctuations, but they cannot
account for the \tsq\ values inferred from several methods of temperature
determination. It is shown that an anomalous broadening of nebular emission lines
appears to have the same sense of correlation as the temperature anomalies,
although a causal link is not obvious.

\end{abstract}


\section{Introduction}

Photoionization models for chemically homogeneous gaseous nebulae of
constant density predict an almost constant temperature; 
consequently, observers often assume a constant temperature to
determine chemical abundances. The spatial temperature inhomogeneities in
\HII\ regions are usually characterized by \tsq\, the mean square
temperature variation in three dimensions.
Recent reviews on the temperature structure of gaseous nebulae are those by
Peimbert (1995, 2002), Esteban (2002), Stasi\'nska (2002), Liu (2002),
Torres-Peimbert \& Peimbert (2003).

{From} uniform-density chemically-homogeneous photoionization models it has been
found that \tsq\ is in the 0.002 to 0.025 range, for objects with
solar or subsolar chemical abundances, with typical values around
0.005  (e.g., Gruenwald \& Viegas 1992, Garnett 1992). Baldwin 
\etal\ (1991, henceforth B91) produced a 
photoionization model for the Orion Nebula, in which grains
and gas are well mixed, that included photoelectric heating
and cooling of the gas by grain ionization and grain collisions respectively;
for this model Peimbert, Storey, \& Torres-Peimbert (1993) found that 
\tsq\ = 0.004.

Alternatively from observations of NGC 346 in the SMC (Peimbert,
Peimbert, \& Ruiz 2000), 30 Doradus (Peimbert 2003), and the
Orion Nebula (Peimbert et al. 1993, Esteban \etal\ 1998, henceforth EPTE98)
it has been found that \tsq\ is in the 0.02 to 0.04 range. The \tsq\
values were derived by combining the temperature derived
from the ratio of the \OIII\ $\lambda\lambda$ 4363, 5007 lines
[$\rm T_{(4363/5007)}$] with the temperature derived from the ratio of the Balmer
continuum to $I(H\beta)$ [$T_{({\rm Bac}/H\beta)}$] and by combining
$T_{(4363/5007)}$ with the temperatures derived from the ratio of
$\lambda$ 5007 to the recombination lines of multiplet 1 of O~{\sc ii}
and the ratio of the of $\lambda\lambda$ 1906, 1909 [C~{\sc iii}] lines to 
the $\lambda$ 4267 C~{\sc ii} recombination line.

The differences between the observed \tsq\ values and those predicted
by models have two profound implications for the study of ionized
gaseous Nebulae: they significantly affect the derived abundances
and they imply the presence of one or several physical mechanisms
not considered by the models.

To study this problem further we decided to obtain an independent  
measurement of \tsq\ based on over a million temperatures determined
across the face of the Orion nebula. The subarcsecond resolution
of the HST makes it ideal for this purpose. Each of these temperatures
represents an average along a given line of sight. Since the
columnar temperatures already represent an average temperature it is
not possible to obtain directly from them the full \tsq\ value. Instead
it is possible to obtain the mean square
temperature variation across the plane of the sky and from this
value it is possible to estimate  \tsq\ . 

In \S\ 2 we discuss the observations. In \S\ 3 and \S\ 4 we present
the method for deriving the columnar temperature ($T_c$) image
together with its analysis. In \S\ 5 we formally define the columnar
temperature, and the mean square temperature variation measured in the
plane of the sky (\tsqa), then determine \tsqa\ from the \OIII\
data. We then estimate \tsq\ based on \tsqa\ and additional data from
the $T_c$ image and \NII\ temperatures. In \S\ 6 we compare our
results with those of previous work.  A new model for the
structure of the Nebula near the ionization front is presented in \S\
7. Finally, the conclusions are presented in \S\ 8. 

\section{The Observations}

The intent of this observational program was to establish accurate \OIII\ 4363~\AA\ and 
5007~\AA\ line ratios across the Orion Nebula using the WFPC2 camera (Holtzman \etal\ 1995) of the Hubble Space Telescope (HST)
and its narrow-band interference filters. This required imaging the nebula in
the filters intended to isolate these lines (F437N and F502N, respectively). Since
the extinction varies at small scales  across the face of the nebula (O'Dell \&\ Yusef-Zadeh 2000), the magnitude of the extinction was determined
on the same size scale as the \OIII\ line observations. This required determination
of the \Ha\ and \Hb\ line ratios, which demanded observations with the F656N and 
F487N filters. Since the former filter allows some emission from the nearby \NII\
doublet to contaminate the \Ha\ signal, it was also necessary to observe with the
F658N filter, which is dominated by the \NII\ 6583~\AA\ line. There was the possibility
of the \Hg\ line contaminating the F437N signal. The anticipation was that this could be
handled by scaling from the \Hb\ observations in F487N.
Each emission line 
filter also passes the nebular continuum radiation. In the case of the emission 
line signal being very strong as compared with the underlying continuum it is
sufficient to measure the visual continuum with the F547M filter and extrapolate
to the wavelength of the emission line filter (O'Dell \&\ Yusef-Zadeh
2000, O'Dell 1998a). However, the 
continuum of the Orion Nebula is quite strong (O'Dell \&\ Hubbard 1965, B91) because the large optical depth of dust particles lying just beyond
the main ionization front (MIF) back-scatter a significant fraction of the
visual starlight reaching it. This back-scatter produces a continuum much stronger than that arising
from atomic processes.  The observations most sensitive to continuum contamination
are those of the intrinsically weak 4363~\AA\ line.  Fortunately, the F469N filter
provided a good measure of the strength of the local continuum because it is
free of strong emission lines when used to observe a low ionization object like the
Orion Nebula. It is known that the He~II 4686~\AA\ line (which the F469N filter
was designed to isolate in high ionization objects) is undetected in the Orion Nebula even in spectroscopic studies 
reaching nearly 10$^{-4}$ the intensity of \Hb\ (B91, EPTE98). Combining the F547M
and F469N observations allowed a linear extrapolation or interpolation of the
wavelength dependence of the continuum flux to the wavelength of each filter.
The filters employed and their pre-launch characteristics (WFPC2 Instrument Handbook, version 6.1
available at http://www.stsci.edu) are 
summarized in Table 1. This table also gives the WFPC2+HST system throughput as recently determined
in orbit.

Because of the host of filters that had to be used and the length of the observing
times necessary for producing an adequate signal in the fainter lines, only one
field of view was covered.  This was selected to include regions that had been
spectroscopically studied recently using high dynamic range detectors (B91 and EPTE98),
in order to allow a precise calibration of the filters,
and to avoid the presence of the bright Trapezium stars. As will be noted in 
\S\ 3.3, the brightest Trapezium star (\thC) was so close (7\arcsec) to the CCD1 detector that
low-level scattered light problems affected the long exposure images with that
detector, rendering them useless for this program. The \OIII\ image is shown in
Figure 1.  The four CCD detectors of the WFPC2 are numbered counter-clockwise,
with the high-angular resolution, but smaller field of view, CCD being designated
as CCD1.

All exposures were made with exactly the same pointing on 2002 January 21, 22, and 26
and at the same gain (electrons/ADU) values.
Multiple exposures of the same duration were made with each filter in order to allow correction
for cosmic ray induced events on the detectors. The individual exposure times and
number of exposures were: F437N(1100 s, 6x), F469N(1100 s, 6x), F487N(350 s, 3x),
F502N(160 s, 3x), F547M(230 s, 3x), F656N(140 s, 2x), F658N(200, 2x). All images
were ``on-the-fly'' calibrated by the Space Telescope Science Institute (STScI)
and then processed using their STSDAS and the National Optical Astronomy Observatories IRAF
\footnote{IRAF is distributed by the National Optical
Astronomy Observatories, which is operated by the Association of
Universities for Research in Astronomy, Inc.\ under cooperative
agreement with the National Science foundation.}
tasks. The final product of the routine processing was a combined image such as
shown in Figure 1.  

\section{Derivation of the \Tc\ Image}

A considerable number of steps were necessary to convert the photometric observations with
WFPC2 into \OIII\ 4363~\AA\ and 5007~\AA\ line ratios and from these line ratios into 
\Tc\ values. In this section we first describe the method of calibration of the
filter properties and other components of the photometric system and their application to
obtaining line ratios. We then describe the method of correcting for interstellar reddening. Finally,
we derive the pixel by pixel values of \Tc\ and estimate the statistical noise of the \Tc\ values.

\subsection{Method of Calibration of the Filters}

The method of calibration adopted was identical to that employed by O'Dell \&\ Doi (1999, henceforth OD99).
However, in the present case we are only interested in the relative line fluxes, so that their
equations for an absolute calibration can be simplified.  The reader interested in absolute calibrations
should refer to OD99, but note that an editing error occurs in their equation (5). The term that reads
$\rm N_{658}t_{656}T^{656}_{N~II}$ should read $\rm N_{658}t_{656}T^{656}_{H\alpha}$, this error does not affect
any of the numerical results in that paper.
The basic approach is that one uses
the WFPC2 observations of spectroscopically well studied regions to determine the calibration
constants, which can then be applied to all the pixels of the image since the filter profile and
system sensitivity does not vary across the image (after the appropriate flat-field corrections have
been applied). The most important calibration constants are determined directly from the observations
of the reference regions and only the secondary constants draw on filter characteristics that were last
determined before launch of the WFPCs. Since the filter and detector characteristics can change with
time, this method of calibration permits accurate determination of the most important characteristics.

Our method of deriving relative calibrations relates N$\rm _{filter}$ (the recorded number of ADU counts per unit 
time) to the characteristics of the source and the filter and camera system through the equation
\begin{equation}
\rm N_{filter}=\sum s_{line} T^{filter}_{line} S_{line} + s_{cont} E_{filter} S^{filter}_{cont},
\end{equation}
\newline 
where we have used the following nomenclature:
\vspace{1cm}
\newline $\rm s_{line}$=system throughput of the telescope-camera-detector system for a
particular line (excluding the filter)
\newline $\rm T^{filter}_{line}$=transmission at a particular line of a filter
\newline $\rm S_{line}$=surface brightness of a particular line in photons $\rm cm^{-2}
 s^{-1} ster^{-1}$
\newline $\rm s_{cont}$=system throughput of the telescope-camera-detector system for the
continuum (excluding the filter)
\newline $\rm E_{filter}=\int_{0}^{\infty} T_{\infty} \mbox{d}\lambda$=effective
width of the filter in \AA 
\newline $\rm S^{filter}_{cont}$=surface brightness of the continuum in photons $\rm cm^
{-2} s^{-1} ster^{-1} \AA^{-1}$.
\vspace{1cm}
\newline

It is important to note that this calibration is done in units of the number of photons. 
When necessary, the resulting numbers can be converted to the relative flux.
In this program multiple lines contributed to the signal from a single filter only in the
case of F656N (\Ha-6563~\AA, \NII-6548~\AA+6583~\AA) and F658N (\NII-6583~\AA, \Ha-6563~\AA).
We considered the contribution of \Hg\ 4340 to the F437N filter, but found it to be negligible.

The reference regions were all taken from B91. They observed with a long slit of east-west orientation
located to the east of \thC. The exact location was incorrectly expressed in B91, but is correctly described
in OD99. B91 gives line ratios and continuum to \Hb\ ratios for 21 samples along their slit. We used the
six samples that overlapped with the CCD4 portion of our images (shown in Figure 1) that included 4363~\AA\ relative intensities. Each 
sample was 2\arcsec x 15.76\arcsec\ in size.

The results of this calibration gave the following equations for the two important line ratios,
\Ha/\Hb\ and 4363/5007.

\begin{equation}
\rm
\frac{S_{H\alpha}}{S_{H\beta}}= 0.443\frac{N_{656}}{N_{487}}\frac{[1 - 0.17\frac{N_{658}}{N_{656}}(1 + 0.19\frac{N_{547}}{N_{658}})]}{(1 - 1.28\frac{N_{469}}{N_{487}})},
\end{equation}
\newline
and
\begin{equation}
\rm
\frac{S_{4363}}{S_{5007}} = 1.90\frac{N_{437}}{N_{502}}\frac{[1 - 0.787\frac{N_{469}}{N_{437}}(74.3\frac{N_{469}}{N_{547}}-0.28)]}{[1 - 1.62\frac{N_{469}}{N_{502}}(33.4\frac{N_{469}}{N_{547}}+0.423)]}.
\end{equation}
\newline

Equation (2) uses a full-image  average value for the color of the continuum and scales the continuum intensity
from the closer continuum filter. Equation (3) derives the color of the continuum at each pixel and then extrapolates or interpolates from the two continuum filters, this step being necessary as the continuum correction is quite
important in the F437N measurements. The continuum typically contributes about 73\%\ of the total signal in
the  F437N filter.

In a similar investigation of planetary nebulae Rubin \etal\ (2002,  henceforth R2002) argued that there should be a small correction
in the F437N filter for contamination in the blueward wing by \Hg\ 4340~\AA\ emission. If one takes the
pre-launch transmission profiles in the Instrument Handbook at face value, this would seem to be
the case. However, it must be expected that the profiles of the filters change as they age in space. These
changes are why we wanted to calibrate our observations within the data set.  One would expect that the 
transmission near the principal lines to not change much with time, since they are located near the peak of
the filters' transmissions, whereas the \Hg\ line falls far out on the wing of the filter curve. The Instrument
Handbook indicates that the CCD sensitivities and throughput of the system have been remarkably constant.
This combination of factors allowed us to empirically search for evidence of \Hg\ contamination. We could 
determine the ratio $\rm s_{437}T^{437}_{4363}/s_{487}T^{487}_{4861}$ with and without a correction for \Hg\ 
contamination. With the correction the ratio was 0.42 and without the correction the ratio was 0.68. The 
Instrument Handbook values indicate an expected value for the ratio of 0.62. This indicates that the correction
to the 4363~\AA\ line surface brightness for \Hg\ is about 10\%.  Since the method of calibration essentially
forces the calibration constants to be such as to give the right intensity ratios for the lines assumed to
be present, this means that the effects of \Hg\ correction will only enter in regions where the \OIII\ line
and the \Hg\ line vary considerably in relative intensity. This is not the case for the area covered by
the present images (B91, EPTE98).

The scatter of results from the six different samples in the calibration region indicates that the uncertainty
of the Balmer line ratios will be 7\%, while that of the \OIII\ line ratios will be about 12\%. Since the 
reddening corrections between 4363~\AA\ and 5007~\AA\ are small, the effects of the absolute uncertainty of
the Balmer line ratios are negligible. A 12\%\ systematic uncertainty of the \OIII\ line ratios would make the derived
\Tc\ values be off systematically by $\pm 350$~K.
This will not significantly affect our search for fine scale variations in
\Tc. As we will see in \S\ 3.3, there is excellent agreement between the electron temperatures derived in B91 and
those derived in this study.

\subsection{Derivation of Extinction Corrections}

The extinction corrections are based on determining the Balmer line ratio for each pixel and then calculating
the total extinction from comparison with the intrinsic ratio. This total extinction was then used to correct
the \OIII\ line ratios. We assumed with EPTE98 that the intrinsic Balmer line flux ratio was 2.91. Using the
same extinction law as EPTE98 means that the extinction corrected ($I$) flux ratios could be calculated from the
observed flux ratios ($F$) by the relation
\begin{equation}
{\rm log}\frac{I_{4363}}{I_{5007}}
={\rm log}\frac{F_{4363}}{F_{5007}} 
+ 0.362~{\rm log}\frac{F_{\rm H\alpha}}{F_{\rm H\beta}} 
-0.168.
\end{equation}
\newline

The extinction values derived from the Balmer line ratios are in good agreement with those of
O'Dell \&\ Yusef-Zadeh (2000), this being the case for both their results obtained from radio continuum to
\Ha\ ratios and those obtained from their Balmer line ratios.

\subsection{Derivation of \Tc\ Images}

The \OIII\ reddening corrected line ratios were then used to derive local values of \Tc.
A convenient formulation of the relation of line ratio and \Tc\ is given in R2002. Their equation (8) can be
restated as 

\begin{equation}
T_{c} = 32,966/({\rm ln} \frac{I_{5007}}{I_{4363}} - 1.701),
\end{equation}
\newline
where we have restated the relation using the common-usage designation of the wavelengths in air (rather
than the more physically meaningful vacuum wavelengths employed by R2002).
Since we are dealing with large numbers that are not usually determined to high precision, we sometimes
will express \Tc\  by the frequent nomenclature \tfour $\equiv$\Tc /10,000.
The resulting image is shown in Figure 2.

Figure 2 shows multiple features. The most common are the ``scars'' that result from stars. 
These continuum sources are often saturated on some of the images used to derive \Tc. Even when
they are not saturated, the formal value of \Tc\ at the star positions is subject to big uncertainties owing to
the multiple subtractions that occur in the derivation. Within CCD1 there are multiple features that are
related to scattered light from \thC\ and the more westward members of the Trapezium. Any apparent fluctuation
in \Tc\ within CCD1 is suspect as being an artifact. There are a few features characterized by very low 
values of \Tc, which are discussed in \S\ 4.3.

This derivation gives for the six sample regions from the B91 study $<$\tfour$>$=0.92$\pm$0.01, while values given in 
B91 for the same samples gives $<$\tfour$>$=0.91$\pm$0.02. A comparison with the EPTE98 study is not possible because
only their position 2 fell within a part of our image that was not useful as it fell within CCD1 and their position 1 was north of our field.

\subsection{Expected statistical noise and its effect on \Tc}

It is important to assess the effect of the expected Poisson noise on the determination of \Tc\ because
of the low signal level in some of the filters. In the region that was used for calibration the total
number of recorded electrons ( 7 times the number of ADU's recorded per pixel) for the key filters were
F437N(1260), F469N(1260), F547M(8400), and F502N(10500).  In the following discussion, we'll assume that
the root mean square deviation of a signal N is N$^{1/2}$. This is what we will mean as the probable error (p.e.).
We will assume that for multiplicative factors the p.e.'s add quadratically.
Unfortunately, one cannot determine the p.e. directly from the observations since the nebula is intrinsically
variable in surface brightness.

An assessment of the p.e. per pixel was done by creating noisy images (using IRAF task {\it mknoise}) of the
appropriate noise level for each of the observationally determined components of equation (3), then determining
the p.e. of the resulting image. The p.e. was 20\%.  In most of the following discussions we 
will assume this uncertainty.

Equation (5) was then used to determined the effect of uncertainties of the \OIII\ line ratio on the derived
value of \Tc. A 20\%\ photometric p.e. produces a p.e. in \Tc\ of 5.7\%.  Throughout this paper
we will present uncertainties as probable errors (p.e.) calculated as the reciprocal of the 
root-mean square deviation of the recorded signal.

\subsection{The Effect of Contamination by weak \Fe\ and \He\ lines}

The assumptions used in \S\ 3.1 for the derivation of Eq. 3 are not fully justified
since there are two weak lines that affect the signal in the more important continuum
filter, F469N.  The \Fe\ line appears at 4702 \AA\ and the \He\ line at 4713 \AA,
where the transmissions are 0.51 and 0.15, respectively. As we will demonstrate in
this section, the effect of these lines is small in determination of \Tc\ in the 
nebula.  

The presence of these weak lines will make the continuum at F469N appear brighter 
than it is.  If we call the enhanced signal in the F469N N$\rm _{OBS}$ and the signal
that would be observed if there were only continuum radiation N$\rm _{CONT}$, then their
ratio will be given by 

\begin{equation}
\rm \frac{N_{OBS}}{N_{CONT}}= 1 + \frac{T^{F469N}_{4702}}{E_{F469N}} \frac{S_{4702}}{S_{CONT,F469N}} + \frac{T^{F469N}_{4713}}{E_{F469N}} \frac{S_{4713}}{S_{CONT,F469N}}.
\end{equation}
\newline

The strength of the continuum is usually measured with respect to the \Hb\ line at
4861 \AA. This was done in B91, where their figure 2 presents a quantity which is
related to the equivalent width (\Whb) by $\lambda$/\Whb. The average value of
their plotted quantity is 9.5, which gives an average \Whb = 510 \AA, which is in
good agreement with earlier filter photometry over a wide range of the nebula
(O'Dell \&\ Hubbard 1965) which showed that \Whb\ is nearly constant in the
inner region of the nebula, then decreases outward. Having values of \Whb\ allows us
to express the observed-quantity ratios of equation (6) as 
S$_{4702}$/S$\rm _{CONT,F469N}$=\Whb S$\rm _{4702}$/S$_{H\beta}$ and
S$_{4713}$/S$\rm _{CONT,F469N}$=\Whb S$\rm _{4713}$/S$_{H\beta}$. Making this 
substitution and inserting the known values allows the restating of equation (6) as

\begin{equation}
\rm \frac{N_{OBS}}{N_{CONT}}= 1 + 14.95 \frac{S_{4702}}{S_{H\beta}} + 4.397 \frac{S_{4713}}{S_{H\beta}}.
\end{equation}
\newline

If the ratio of line intensities with respect to \Hb\ are constant across the
region observed, then there is no net effect, since they will be constant in the 
calibration region and their affects will cancel out.  However, if the line ratios
vary, there will be an effect on the derived line ratios. Anticipating that this
effect is small, we address only the effect on equation (3).  Since the correction
term in the denominator of that equation is only about 0.015, we can demonstrate adequately
the effect of varying contaminating line strengths by considering only the terms
in the numerator.  If (N$\rm _{OBS}$ / N$\rm _{CONT}$)$\rm _{CALIB}$ is the value
of the ratio of these quantities in the region used for calibration, we can express
the local effect by multiplying the N$\rm {469}$ rates by
(N$\rm _{OBS}$ / N$\rm _{CONT}$)/ (N$\rm _{OBS}$ / N$\rm _{CONT}$)$\rm _{CALIB}$.
The global average values (taken from a section of CCD2 free of the peculiar 
effects due to shocks that are discussed in \S\ 4.2 are N$\rm {437}$/N$\rm _{502}$=
0.0094, N$\rm {469}$/N$\rm _{437}$=1.04, and N$\rm {469}$/N$\rm _{547}$=0.016.
Inserting these values into the numerator of equation (3) and applying a constant
correction of 0.015 for the subtractive term in the denominator, gives

\begin{equation}
\rm \frac{S_{4363}}{S_{5007}} = 0.0181\{
1 - 0.819 (\frac{N_{OBS}}{N_{CONT}}) (\frac{N_{CONT}}{N_{OBS}})_{CALIB}[1.19 (\frac{N_{OBS}}{N_{CONT}}) (\frac{N_{CONT}}{N_{OBS}})_{CALIB} -0.28]\}.
\end{equation}
\newline

The weak contaminating lines are seen only at high spectral resolution.  In the
study of Osterbrock, Tran,  \&\ Veilleux (1992), fluxes relative
to \Hb\ were 0.00245 and 0.00686 for 4702 and 4713 respectively (for an east-west region 47\arcsec\ long centered 58\arcsec\ north of \thC). EPTE98 found ratios of 0.00179 
and 0.00645 for a region 45\arcsec\ north of \thC\ and 0.00186 and 0.00621 for a region25\arcsec\ south and 10\arcsec\ west of \thC.  These three samples would indicate
values of N$\rm _{OBS}$ / N$\rm _{CONT}$ of 1.066, 1.056, and 1.055 respectively.
The contaminating lines were not measured in B91 due to their lower spectra 
resolution.  

There is another set of data that provide even more
information. Co-authors Antonio and Manuel Peimbert together with Mar\'{\i}a Teresa Ruiz 
obtained a long slit spectrum with the VLT3 telescope
(Melipal) and the FORS1 spectrograph at the Paranal observatory in
Chile on 2002 September 10. The slit was oriented at a position
angle of 340\arcdeg\ and passed through the feature called 
HH~202-S in O'Dell \etal 1997 (henceforth O97)and designated in Figure 2.  To estimate the
intensity of the \Fe\ 4702 \AA\ line, they measured the intrinsically
stronger \Fe\ line at 4658 \AA\ and we scaled it to the 4702 intensity
by using the line ratio of 0.32 from EPTE98.  Likewise, they also
measured the strength of the 4713 \AA\ feature, a blend of [Ar~{\sc
iv}] line at 4711 \AA\ and the \He\ at 4713 \AA\, the [Ar~{\sc iv}] line contributing
about 15\% of the blend intensity. The average of a 2\arcmin\
section, which is close to the east-west calibration samples from B91,
gave ratios relative to \Hb\ of 0.0061 and 0.0060 for 4702 and 4713
respectively, giving a value of N$\rm _{OBS}$ / N$\rm
_{CONT}$=1.0551. The scatter of this value, taken from six 20\arcsec\
subsamples was $\pm$0.0064.  A similar scatter was found from an
analysis of both 20\arcsec\ and 2\arcsec\ samples.

Recalling that the effect of this 5\%\ level contamination cancels out in our
calibration procedure leaves us only to consider the effect of the variation
in the value of \Tc\ due to the varying level of contamination. The variations 
in contamination indicated by the Paranal data indicate that the corresponding
variations in the derived \Tc\ is +80 and -120 K.

\section{Analysis of the \Tc\ Image}

The ``temperature image'' can be used to look for variations in \Tc\ at large and fine scales. In this section
we describe the results of a search for systematic changes, and explain the regions of apparently very low and 
high values of \Tc . We also present the evidence for small scale changes, and give the results of a 
Fourier analysis of the temperature fluctuations.

\subsection{Global Variation of \Tc\ with Distance from \thC}

We looked for variations of \Tc\ with distance from the dominant ionizing star \thC. Nine samples of 10\arcsec
x10\arcsec\ were taken from each of CCD2-4. In each case there were no scars from bright stars and none of the
low \Tc\ regions discussed in the next section. The average gave $<$\Tc$>=$8890$\pm$496 K.  No statistically 
significant correlation was found when trying to fit a linear relation to the observed values of
\Tc\ and distance from \thC.  We conclude that there are real variations in \Tc\
across the region sampled in this study, but there is no obvious relation to the position in the nebula.
The scatter of 496 K is significant since the scatter due to Poisson statistics within each sample would
be only 5 K and we estimated in \S\ 3.5 that the scatter due to variations in the contaminating weak 
lines would be about $\pm$100 K.

\subsection{Anomalous ``Low \Tc '' Regions}

There are several regions within CCD2-CCD3-CCD4 that have anomalously low values of the derived \Tc , as seen
in Figure 2.  These all fall within the province of known Herbig-Haro (HH) objects HH~202 (Cant\'o \etal\ 1980, 
O97)
and HH~269 (Walter \etal\ 1995,  Bally, O'Dell, \&\ McCaughrean 2000).  The former is probably the result of a pre-Main Sequence star's blueshifted jet striking the
foreground veil of neutral material(O97). The latter is probably the result of blueshifted jet outflow from
a pre-Main Sequence star that lies behind the MIF and has yet to be identified (O'Dell \&\ Doi 2003).
It is almost certain that HH~269 has the same source as that producing the series of shocks extending to the
east and designated as HH~529 (Bally, O'Dell, \&\ McCaughrean 2000) and it is likely (O'Dell \&\ Doi 2003) that this is also the source for
the jets producing HH~202 and the double object HH~203+HH~204 that lies to the southeast, near \thtwo.

These anomalously low \Tc\ features are probably not real, rather, they are the result of contamination
of the F469N filter.
Our method of calculation of the line ratios assumed that the F469N filter was
free of any emission lines, i.e. it was used as a measure of the continuum. In all of the nebula for which
spectra exist this assumption is good since the star producing most of the ionization of the nebula
(\thC) is too cool to doubly ionize helium and there are only the two weak contaminating lines (c.f. \S\ 3.5).  Hartigan (1999) has shown that if a fast jet
forms a bow shock in material that is already photoionized then one can expect He~II to be present.
He used ground based images to look for evidence for emission in the He~II 4686 \AA\ line. The only region where
he thought that this line was present was in HH~202, but with his low resolution images he could not
conclude that 4686 \AA\ emission was present.  The present high resolution images show that there are 
\Tc\ anomalies at several of the shocks known to exist in this field of view (but certainly not in all the
known shocks).  The anomalous \Tc\ feature to the southwest from HH~202-S corresponds to a faint shock visible
only in \Ha\ and \OIII, but not in \NII\ (O97). 

The Paranal spectra described in \S\ 3.5 were centered on HH~202-S.  The spectra show no 4686 \AA\ emission.
However, there is a considerable enhancement of \Fe\ emission. In a sample centered on HH~202-S,
the flux ratio of 4658 \AA\ to \Hb\ was 0.031, which predicts a 4702 to \Hb\ ratio of 0.010 (using the flux ratio from EPTE98). Together with a
normal 4713 contribution, this would raise the ratio N$\rm _{OBS}$ / N$\rm _{CONT}$ to 1.18, which would 
artificially reduce the 4363 ratio by a factor of 4.6, and decrease the derived value of \Tc\ to
6300 K.  Since the Paranal sample on HH~202-S was 20\arcsec\ long and the object is much smaller, the 
contamination would be much larger on the object itself and the artificially derived \Tc\ correspondingly
lower.  Direct confirmation of contamination by 4702 was obtained by high velocity resolution spectra obtained by co-author O'Dell
and Will Henney with the Manchester Echelle Spectrometer (Meaburn, \etal\ 1984) on the 2.1-m telescope
at San Pedro Martir observatory on 2002 October 23, where one sees significant enhancement  of that line,
peaking at the optically brightest portion of HH~202-S, with no indication of 4686 \AA\ emission.

This means that enhanced \Fe\ emission is a common feature of the anomalously low \Tc\ spots seen in Figure 2,
rendering useless the \Tc\ values there. Why \Fe\ is enhanced in these shocks is in itself an interesting
problem that should be addressed.

\subsection{Spurious ``High \Tc ''values in this study}

Several proplyds occur within our sample, the brightest within CCD2-4 are 141-301 and
182-413.  These proplyds and the fainter, less well resolved ones all show enhanced values
of \Tc.  In this case the anomaly is probably due to the mis-application of equation (5),
which is applicable only in the limit of electron densities (\Ne) low as compared with
the critical density (\Ncrit) for \OIII.  \Ncrit\ is the density at which collisional
de-excitation of the upper state giving rise to an emission line is equal in probability
to a spontaneous emission. This is only of concern for the 5007 \AA\ line as its \Ncrit\
value is much lower than that for the 4363 \AA\ line. Rubin \etal\ (2003, henceforth R2003) give
\Ncrit (5007 \AA) $\simeq$ 6.4x10$^{5}$ \cmq, which is much higher than densities
encountered in the nebula, but comparable to characteristic values for the proplyds
(O'Dell 1998a).  This means that in the local ionization fronts of the proplyds  the 
5007 \AA\ emission is being suppressed by collisional de-excitation and the \Tc\
value derived from equation (5) will be an upper limit. This same interpretation is
applied by R2003 to their high \OIII\ \Tc\ for the proplyd 159-350, which lies in the
center of our CCD1 image and fell in their long slit spectrum. In their models of the proplyds
close to \thC\ Henney \etal\ (2002) argue that the formal temperatures derived without consideration
of collisional de-excitation would be about 20,000 K. If the densities of the more distant 
proplyds are much lower, as indicated by their surface brightnesses, then the effects of collisional
de-excitation are less and some of the temperature increase we find is real.

It is not clear if the similar local increase in \Tc\ at the jet feature just south of the
dark rim (O'Dell \&\ Yusef-Zadeh 2000) in the southeast corner of CCD4 is due to this
same effect. The jet has a known high radial velocity (O'Dell \etal\ 1997) and proper
motion (O'Dell \&\ Doi 2003), so it may be that \Tc\ is enhanced where this flow shocks
against ambient nebular gas.

\subsection{Small Scale Variations in \Tc }

Figure 2 indicates that there are variations in \Tc\ even though we've shown in \S\ 4.1
that these variations are not a function of distance from \thC.

The smaller scale changes have been illuminated by looking at the three sample regions
depicted in Figure 2, these samples being selected to be free of the [Fe~III] contamination.
The results are shown in Figure 3. In each case the range of temperature is $\pm$ 1000 K.
For reference, random noise at the level of 5.7 \%\ probable error is shown, this being the
level expected for the bright portions. The southwest sample is about one fourth the surface
brightness of the other samples. The noise level was brought down to that of the other samples
by taking samples of 2x2 pixels in this region.
Examination of Figure 3 shows that 
the face of the nebula is mottled with small scale variations of \Tc\ with angular
dimensions of about 10\arcsec\ and amplitudes of about 400 K. 

\subsection{Fourier Analysis of two regions of the \Tc\ image}

In order to gain a more quantitative evaluation of the data, we performed a Fourier analysis of two regions of the \tfour\ image. The first was a 400x400 pixel section in the northwest portion of CCD2 and the second was
250x250 pixels in the northeast section of CCD4, both being free of camera scars and the local
apparent \Tc\ drops that are due to shocks, as discussed in \S\ 4.2.  The analysis was done using 
STSDAS tasks with IRAF. The procedure was the same for each sample.  The first step was to subtract the
average value of the sample from each pixel and then to make a power spectrum of the resulting image,
using the task {\it powerspec}. A radial profile was taken of the power spectrum image using the center
as the origin. The result was a profile that dropped down to a constant value at large 
spatial frequency values and represents the noise level due to the Poisson noise in the data. This constant
value was then subtracted from the radial profile of the power spectrum.  The resulting noise-subtracted
power spectrum was then converted to the physically more meaningful length space by multiplying the power
spectrum in spatial frequency by the square of the frequency and binning the result in the spatial
dimension. The results are shown in Figure 4 and Figure 5. These plots show that there is a large amount
of power at the finest scales (sub-arcsec) but that power is found at all scales. This analysis is similar
to that of Vannier \etal\ (2001) of H$\rm _{2}$ emission arising near the imbedded BN-KL sources to the north of CCD4.

\section{Values of \tsqa (O$^{++}$) and estimates of \tsq (O$^{++}$) and \tsq 
(H$^{+}$)}

The aims of this section are to estimate representative values of
the temperature variations [\tsq (O$^{++}$), \tsq (O$^{+}$), and \tsq
(H$^{+}$)] in the Orion Nebula in order to compare them with 
predictions of photoionization models and to be able to study the
physical conditions of the nebula including its chemical
composition. We proceed in three steps: 1.) use of the
$T_c$ map obtained in this paper to determine \tsqa (O$^{++}$)
across the plane of the sky (this is the first detailed 2-dimensional
set of temperatures across of the Orion Nebula), 2.) estimate 
the 3-dimensional value of \tsq (O$^{++}$); 3.) estimate 
\tsq (H$^{+}$) including both the O$^+$ and the O$^{++}$
regions.

\subsection{Values of \tsqa (O$^{++}$) measured in the plane of the sky}

In order to obtain the observed mean square temperature variation in the plane
of the sky, \tsqa (O$^{++}$), we will define a set of equations.
The average temperature, $T_0(X^{+i})$, and the mean square
temperature fluctuation, $t^2(X^{+i})$, are given by
\begin{equation}
T_0(X^{+i}) = 
\frac{\int T_e({\bf r}) N_e({\bf r}) N(X^{+i};{\bf r}) dV}
{\int N_e({\bf r}) N(X^{+i};{\bf r}) dV},
\end{equation}
and\begin{equation}
t^2(X^{+i}) = 
\frac{\int (T_e - T_0(X^{+i}))^2 N_e N(X^{+i}) dV}
{T_0^2 \int N_e N(X^{+i}) dV},
\end{equation}
respectively, where $T_e$ and $N_e$ are the local electron temperature and density, $N(X^{+i})$
is the local ion density corresponding to the observed emission line,
and $V$ is the observed volume (Peimbert 1967).  By recognizing that
the observations correspond to averages over a given line of sight, it
becomes necessary to define the columnar temperature, \Tc, as
\begin{equation}
T_c(X^{+i}; \alpha, \delta) = 
\frac{\int  T_e N_e N(X^{+i}) dl}{\int N_e N(X^{+i}) dl},
\end{equation}
where $\alpha$ is the right ascension and $\delta$ is the declination
of a given line of sight (corresponding to a given pixel). Then, the average temperature can be written
as
\begin{equation}
T_0(X^{+i}) = 
\frac{\int T_c(X^{+i}; \alpha, \delta)\int N_e N(X^{+i}) dl dA}
{\int\int N_e N(X^{+i}) dl dA},
\end{equation}
and the mean square temperature variation over the plane of the sky,
\tsqa, can be defined as:
\begin{equation}
t^2_A(X^{+i}) = 
\frac{\int (T_c(X^{+i})- T_0(X^{+i}))^2 \int N_e N(X^{+i}) dl dA}
{T_0(X^{+i})^2 \int \int N_e N(X^{+i}) dl dA}.
\end{equation}

In Table 2 we list a set of \tsqa (O$^{++}$) values obtained from
equation (13) and from the $T_c(\alpha, \delta)$ map presented in
Figure 2.  To determine \tsqa (O$^{++}$) we have excluded several
features in order to diminish the noise of the samples: the pixels
with very low flux, where we expect the temperature deviations to be
mostly due to errors and not to physical inhomogeneities; the shocks,
where the destruction of dust could be enhancing the \Fe\ 4702 line
and thus producing spurious temperatures; the proplyds, whose density
is not well represented by equation (5); the seams between the CCDs,
where there are border effects. Keeping in mind these exclusions we
defined the six regions from Table 2 as follows: regions 1 and 2
include most of the pixels of the SE and SW regions as defined in
Figure 3; region 3 includes most of the pixels in a 54" $\times$ 73"
region in the eastern part of CCD4; 
region 4 is the largest sample, it
includes most of the pixels from CCD2, CCD3, and CCD4; finally region
5 includes approximately 72\% of the pixels of region 4 where, in
order to enhance the S/N ratio, the southern 10\% of CCD2, the
southwestern 50\% of CCD3, and the western 25\% of CCD4 have been
removed, the regions removed are the ones of relatively low emission
measure. Although the signal in CCD1 was high, the presence of scattered 
light artifacts (c.f. \S\ 2) prevents its use in our analysis.

Column 2, of Table 2, presents the number of independent temperatures
used in the region; column 3 presents the average temperature,
weighted by $\int N({\rm O}^{++}) \times N_e dl$ (defined in equation
(12) ) ; while column 4 presents the measured \tsqa(O$^{++}$) for each
region.

The \tsqa\ of regions 1, 2, and 3 is in the 0.005--0.016 range.  The
differences in \tsqa\ are real since they are larger than the expected
observational errors. This could have 2 explanations: a) region 3 is
more homogeneous than regions 1 and 2; b) another possibility is that
\tsq\ is similar in the 3 regions, and the variations along each line
of sight in region 3 are larger, while there are not so many
variations along the lines of sight in regions 1 and 2, thus giving a
set of more homogeneous values of $T_c$ for region 3 than for regions
1 and 2.

We added most of the pixels from CCD2, CCD3, and CCD4 to generate
region 4.  Figure 6 presents a histogram of the 1.5 million
temperatures included in region 4. The mean square temperature
deviation obtained from the histogram is $\sigma(T)^2/T^2 = 0.018$;
which is not \tsqa.  It is not possible to derive directly \tsqa\ from
the histogram since each temperature has to be weighted by the
corresponding emission measure (see equation (13)). After weighting each
\Tc\ value for the emission measure we obtain a $t^2_A({\rm O}^{++})=
0.0096$ for region 4.

We still have to subtract a value associated with the noise.  The
average p.e. in $T_c({\rm O}^{++})$ for the whole sample is 5.7 \%, 
however this value is for an average pixel, while the measured
$t^2_A({\rm O}^{++})$ gives more weight to the brightest pixels, the
ones with the smaller errors.  By considering that the average pixel
is 1.89 times fainter than the average pixel weighted by emission
measure, we expect the representative p.e. to be 4.2 \%, which
corresponds to an error on \tsqa\ of 0.0017, that is a $t^2_A({\rm
O}^{++})= 0.0079$.

If we take out the noisiest parts of region 4 (the outermost parts), we
have region 5. Here the measured $t^2_A({\rm O}^{++})= 0.0084$. For
region 5 we expect the noise to be about 2/3 the noise of the entire
region, that is a p.e. of 3.8 \%\ per pixel.  In region 5 the average pixel is
1.31 times fainter than the average pixel weighted by emission
measure; thus we expect the representative p.e. to be 3.3\%, which
corresponds to an error on \tsqa\ of 0.0011, that is a $t^2_A({\rm
O}^{++})= 0.0073$.

We expect the \tsqa\ result for region 5 to be more accurate; however, by
excluding the outermost parts we seem to be ignoring a fraction of the
object which is intrinsically cooler. Therefore we will adopt the value
derived from region 4, $t^2_A({\rm O}^{++}) = 0.0079$, as
representative for the whole object.

This value is similar to the one obtained by R2003 where they find
$t^2_A$ values of 0.0095, for their slit 1, and 0.0104, for their slit
2. Unfortunately, their slits 1 and 2 are located in our unusable CCD1,
which precludes a direct comparison.

\subsection{Relation of $t^2_A$(O$^{++}$) measured in the plane of the
sky to its 3-dimensional value, $t^2$(O$^{++}$)}

Up to this point we have presented the results in terms of $t^2_A$(O$^{++}$), but
what is of physical interest are the variations in three dimensions \tsq($X^{+i}$).
In order to obtain the total \tsq($X^{+i}$) value we need to consider the
variations along the line of sight. It can be shown that the relevant
equation is:
\begin{equation}
t^2(X^{+i}) = t^2_A(X^{+i}) + \left<t^2_c(X^{+i}; \alpha, \delta)\right>,
\end{equation}
where $t^2_c$, the variation along a given line of sight, is given by:
\begin{equation}
t^2_c(X^{+i}; \alpha, \delta) = \frac{\int N_e N(X^{+i}) (T_e - T_c)^2
dl} {T_c^2 \int N_e N(X^{+i}) dl},
\end{equation}
and the average over all lines of sight is given by:
\begin{equation}
\left<t^2_c(X^{+i}; \alpha, \delta)\right> =
\frac{\int t^2_c T^2_c \int N_e N(X^{+i}) dl dA}
{T_0^2 \int \int N_e N(X^{+i}) dl dA}.
\end{equation}

There are two estimates that we can make of $t^2_c$: the first based
on geometrical considerations and the information contained in
$t^2_A(X^{+i})$; and the second based on the ratio of $X^{+i}$
collisionally excited lines to $X^{+i-1}$ recombination lines.

{From} geometrical considerations one would expect the inhomogeneities
of one additional dimension, $t^2_c$, to be at least half as large as
the inhomogeneities of two dimensions, $t^2_A$; this works for large
scale inhomogeneities. The power included in $t^2_c$ increases in the presence of small scale
inhomogeneities. This is because the 
many independent thermal elements along each line of
sight will be masked in the averaging process, greatly reducing
the total $t^2_A$ compared to $t^2$. Therefore, $t^2_c$ would be
expected to be much larger than $t^2_A$. From the previous
considerations, we estimate that $0.5 \la t^2_c/t^2_A \la 2.0$; this
result is very model dependent which is reflected by the large range
in the estimated $t^2_c/t^2_A$ value. This estimate together with the
$t^2_A$(O$^{++}$) value derived in the previous subsection yield
$0.004 \la t^2_c({\rm O}^{++}) \la 0.016$.

{From} the ratio of O~{\sc ii} recombination lines
to \OIII\ lines in two regions of the Orion Nebula EPTE98 find that
$t^2$(O$^{++}) = 0.023 \pm 0.005$. This $t^2$ value is representative
of a very small fraction of the nebula, so we expect it to be a lower
bound to the total $t^2$(O$^{++}$). At the same time it must be bigger
than $t^2_c$(O$^{++}$) because it represents an average over more than
one line of sight. This argument, together with the $t^2_A$(O$^{++}$) value
estimated in the previous subsection, yields $0.010 \la t^2_c({\rm
O}^{++}) \la 0.028$.

By combining both estimates we find $t^2_c = 0.013 \pm 0.005$,
which implies that there is power in small scale
inhomogeneities.
Our prefered value of $t^2_c$ together with the $t^2_A$ measured in
the previous subsection and equation (14) gives us a $t^2({\rm
O}^{++})=0.021 \pm 0.005$ value.

\subsection{An estimate of the temperature variations for the whole \HII\ 
region}

We are
interested in the global $t^2$ value that we will define as:
\begin{equation}
t^2({\rm H\, {\scriptstyle II}}) = t^2({\rm H}^+)
\end{equation}
in order to derive the physical condtions in the \HII\ region.

In an \HII\ region, like the Orion Nebula, where oxygen is only once
or twice ionized, it is useful to define a value $\gamma$, that
characterizes the degree of oxygen ionization, as:
\begin{equation}
\gamma =
{\int N_e N({\rm O}^{++}) dl dA \over 
\int \left[ N_e N({\rm O}^+) + N_e N({\rm O}^{++}) \right] dl dA};
\end{equation}
for such an \HII\ region \tsq(\HII) is given by (Peimbert, Peimbert,
\& Luridiana 2002):
\begin{equation}
t^2({\rm H\, {\scriptstyle II}}) = 
\gamma t^2({\rm O}^{++}) {T_0({\rm O}^{++})^2 \over T_0({\rm H}^+)^2} + 
(1 - \gamma) t^2({\rm O}^+) {T_0({\rm O}^+)^2 \over T_0({\rm H}^+)^2} + 
\gamma (1 - \gamma) {[T_0({\rm O}^+) - T_0({\rm O}^{++})]^2 \over T_0({\rm 
H}^+)^2},
\end{equation}
the first term corresponding to the temperature variations in the
O$^{++}$ zone, the second term corresponds to the variations in the
O$^+$ zone, and the third term corresponds to the thermal variations
introduced by having two zones with different average temperatures.

{From} the observations of Peimbert \& Torres-Peimbert (1977) and EPTE98
it follows that $\gamma$ decreases from about 0.85 near the center of
the nebula to about 0.15 in the outer regions. We will adopt a value
of $\gamma = 0.6$ as representative of the whole object.

It is necessary to estimate \tsq(O$^+$). Since $T({\rm O}^+)$ is
rarely measured we will use $T({\rm N}^+)$ as a surrogate since
we expect both temperatures to be similar as they arise from the same
narrow \hplus + \heneutral\ zone.
R2003 present values of $t^2_A({\rm N}^+)$; their measured
values are in very good agreement with their measured values of
$t^2_A({\rm O}^{++})$; therefore we will assume $t^2_A({\rm O}^+) =
t^2_A({\rm O}^{++})$. It is impossible, however, to estimate
$t^2_c({\rm O}^+)$ from any set of observations available. From the
geometrical arguments presented in the previous section and the
similarities between the $t^2_A({\rm O}^{++})$ and $t^2_A({\rm O}^+)$
presented by R2003 we will assume also that $t^2_c({\rm O}^+) = t^2_c({\rm
O}^{++})$, and consequently that $t^2({\rm O}^+) = t^2({\rm O}^{++})$.

To compute the final term in equation (19) we need to estimate the
differences in temperatures between the O$^+$ and O$^{++}$ zones
(again we will use $T({\rm N}^+)$ as a surrogate). Our observations do not yield
$T({\rm N}^+)$ values, therefore
we are forced to look for them in the literature. We are
more interested in finding $T({\rm N}^+) - T({\rm O}^{++})$ than
$T({\rm N}^+)$ to avoid possible biases, since the calibration is
uncertain by a few hundred degrees (see final paragraphs of sections
3.1 and 3.5); an additional advantage of such comparison is that
papers in the literature cover regions considerably smaller than the
one presented in this paper. EPTE98 have measured $T({\rm
N}^+)=10200$~K and $T({\rm O}^{++})=8300$~K; while R2003, for their
slits 1 and 2 and assuming $N_e= 5000\, {\rm cm}^{-3}$, obtain
$T({\rm N}^+)=10050$~K and $T({\rm O}^{++})=8200$~K. 
By averaging the innermost 9 regions of B91, we find $T({\rm
N}^+)=10200$~K and $T({\rm O}^{++})=9050$~K. These three results give
us an average $T({\rm N}^+) - T({\rm O}^{++}) = 1650$~K.
As mentioned before, we have adopted a $\gamma = 0.6$, giving a
$T_0({\rm H}^+)$ approximately 700~K higher than $T({\rm O}^{++})$.

The inputs presented above yield values of 0.011,
0.010, and 0.007 for the three terms in equation (19). Therefore the
total magnitude of the temperature fluctuations becomes $t^2({\rm H\,
{\scriptstyle II}}) = 0.028 \pm 0.006$.  Notice that for values of
$\gamma$ in the 0.3 to 0.7 range the change in total \tsq\ is very
small since $t^2({\rm O}^+) = t^2({\rm O}^{++})$, thus changes in
$\gamma$ will make very small changes in the sum of the first 2 terms
of equation (19), while the third term will change by less than 15\% for
that range of $\gamma$.

The total $t^2({\rm H\,{\scriptstyle II}})$ value derived in this
paper is considerably larger than the 0.004 value derived by Peimbert
et al. (1993) from the one dimensional model of the Orion Nebula
presented by B91.

\section{Comparison with previous work}

The type of temperature variations reported here have not been seen before because previous slit
spectra determinations usually employed integrations over long
areas of the nebula. The best slit spectrum sub-sampling until R2003 is in B91,
where the sample sizes were as short as 16\arcsec\ and one could see
sample to sample changes in \Tc\ of about 300 K (B91, Table 5). 
Their changes were comparable to the uncertainties in their
derivations of \Tc\ from the \OIII\ line ratios, so this variation did not attract
attention. However, the B91 variations are consistent with the small
scale variations we see in our Figure 3.

A much more valuable study is that of R2003, where they employed the
HST's STIS (Space Telescope Imaging Spectrometer) in a long slit mode
to determine \Tc\ in samples of 0.5\arcsec x 0.5\arcsec\ in both
\OIII\ and \NII. They presented results for about 400 samples from
each of four slit settings, two within our CCD1 field and the others
near or off the southeast corner of our CCD2 field. It is possible to
use this data to estimate a lower limit to $t^2$(H~{\sc ii}) using the
same procedure as in the previous section.

{From} their slits 1 and 2, those included in our map, it is found
that $T_0({\rm O}^{++})=8200$~K and $T_0({\rm N}^+)=10050$~K; notice
that these $T_0$'s corresponds to two rectangular columns of the nebula and not to a
large volume. From these slits we also find $t^2_A({\rm O}^{++})=0.010$
and $t^2_A({\rm N}^+)=0.009$.

{From} R2003 observations it is not possible to obtain direct estimates
of the O$^{++}$ and N$^{+}$ temperature fluctuations along any given line of
sight; so we will use $t^2_A({\rm O}^{++})$ and $t^2_A({\rm N}^+)$ as
lower bounds to $t^2({\rm O}^{++})$ and $t^2({\rm N}^+)$.

To estimate the two zone temperature term, the term due to the
difference between $T_0({\rm O}^{++})$ and $T_0({\rm N}^+)$ (see
equation (19) , we will adopt $\gamma=0.7$. This, together with
$T_0({\rm O}^{++})$ and $T_0({\rm N}^+)$, yields a contribution of
0.009.

Finally, adding these 3 terms, we obtain $t^2$(H~{\sc ii})$ \ge 0.019$.
This lower limit to $t^2$(H~{\sc ii}) is consistent with our
determination and with the $t^2$ determination by EPTE98. Moreover it
is already a factor of 5 higher than the values predicted by
one-dimensional photoionization models of the Orion Nebula, like the
one by B91.

\section{Discussion}

In this section we will review the large scale structure of the Orion Nebula (\S\ 7.1), present
a conundrum arising from the presence of fine scale structure in images (\S\ 7.2), demonstrate that
high resolution spectra and line ratio changes argue for a three dimensional fine scale structure 
(\S\ 7.3), argue that the ionization front is advancing into a highly clumpy PDR (\S\ 7.4), 
show how these clumps can lead to low \Te\ shadows within the ionized layer that can 
help to explain the \tsq\ phenomenon in the region of low ionization regions (\S\ 7.5), 
and present a preliminary discussion of the effect of these clumps on the anomalous line-broadening in Orion (\S\ 7.6). 

\subsection{The global structure of the Orion Nebula}

The accepted model for the Orion Nebula is that of a blister. 
The visible nebula is the result of a thin layer of photoionized gas on the front of a giant
molecular cloud, with a dense photon-dominated region (PDR) (van der Werf \etal\ 1996), also known (Young Owl \etal\ 2000) as a photodissociation region, on the neutral side of the 
ionization front. Gas is accelerated away from this ionization. The dominant ionizing
star is \thC\  which lies at about 0.25 pc in front of the ionization front. In the foreground is an
irregular veil of neutral material that produces most of the extinction (O'Dell 2002). This model is reviewed in detail in a recent review article (O'Dell 2001b, henceforth OD2001.  
The basic physics is discussed in a review paper by Ferland (2001).

The shape of the ionization front has been derived (Wen \&\ O'Dell 1995, henceforth WO95) using a 
basic process originally proposed by Gary Ferland in B91. This draws on the fact that
at a substellar point, the surface brightness in a recombination line (for example \Ha) is
directly proportional to the local flux of ionizing photons. WO95 solved the more general problem,
where the line of sight to a point in the nebula does not pass through the ionizing star. They
showed that the ionization front is an irregular concave form, with an escarpment that explains
the Bright Bar feature running near \thtwo\ and has a dominant hill to the southwest of \thC,
where the Orion-S radio and infrared sources are located.  O'Dell \&\ Doi (2003) have shown that
the latter region is the source of multiple optical and radio outflows and probably contains
a secondary source of star formation. This model was derived using a resolution of 2\arcsec.
Because of the progressive nature of the calculation of the shape of the surface, it is most accurate
near the direction of \thC.  

The only diagnostics that we have along the line of sight into the nebula
are the increasingly blue-shifted emission from the higher ionization states that are found
further from the MIF and forbidden line doublet intensities that indicate a decreasing
density with increasing distance from the MIF (OD2001). These few tests that we have are consistent
with the model.  

A detailed hydrodynamic model of the Orion Nebula has not been calculated,  however, it is 
expected that the high density in the PDR region is the result of a shock moving into the molecular cloud and that
the ionization front is slowly progressing into the neutral region.  

\subsection{A quandary presented by the appearance of fine scale structure in the Orion Nebula}

The HST images of the Orion Nebula are filled with detail, especially in the inner regions.
At first blush, this appears to present 
a fundamental quandary if the standard model for the nebula is correct.
One can derive a formal thickness (L) for the ionized layer, using the
assumption that the density is constant and knowing the emissivity in the recombination
line being observed. WO95 show that this value is characteristically
about 0.13 pc for the central part of the nebula, which corresponds to an angle 
of 58\arcsec\ if the distance to the nebula is 460 pc (Bally, O'Dell, \&\ McCaughrean 2000).
They also show that if the density has an exponential decrease, its scale-height will be
L/2 (29\arcsec) and the emissivity would decrease with a scale-height of L/4 (15\arcsec).
This means that in the standard model one should not be able to see variations
in the structure of the nebula with size scales smaller than about 15\arcsec. 

One certainly see lots of fine-scale (less than 15\arcsec) structure in ground-based and HST images. 
Some of this can be attributed to variations in the amount of extinction (O'Dell
\&\ Yusef-Zadeh 2000), most of which is caused by the foreground neutral veil of material (OD2001).
However, the extinction corrected images in \Ha\ show significant structure right down to
the limit of the extinction corrections, which was about 1.7\arcsec. This is
shown in Figure 7,
which is a profile of the extinction corrected \Ha\ image of the nebula (O'Dell \&\ Yusef-Zadeh 2000)
derived  with a resolution of 1.7\arcsec\ and taken across the velocity sample region 
designated in Figure 1.  

There are even smaller features, which appear to be a continuation of the structure, as shown in
the HST images, while the smallest features 
are due to the outflows from young stellar objects and the shocks they create in the low density ambient gas (O'Dell \&\ Doi 2003, OD2001)
and are seen in the HST images to be sub-arcsecond in extent.

The presence of fine scale structure associated with the main emitting layer of ionized gas presents
a fundamental challenge to the accepted model of the nebula. The structure is there, for we see it,
but if the ionized gas was simply a smooth flow of gas away from a homogeneous ionization front, we
should not see any small scale structure.
We demonstrate in \S\ 7.5 and \S\ 7.6, that the
resolution of the quandary lies in the emitting layer having a much more complex structure than assumed 
in published models.

\subsection{Substructure in the main emitting layer revealed by fine scale velocity variations and line ratios}

There are two additional lines of evidence indicating that small scale structure exists within the ionized layer,
these being the structure seen in the velocities and fine scale variations in the ionization. Both
argue for structure down to at least a few arcseconds.

The internal velocities within the Orion Nebula have been the subject of investigation for nearly
a century, beginning with a Fabry-Perot study by Fabry (Buisson, Fabry, \&\ Bourget 1914), an extensive mapping using 
photographic detectors (Wilson \etal\ 1959), and a series of studies in various emission lines using
CCD detectors ([O~III], Casta\~neda 1988; [O~II], Jones 1992; [O~I], O'Dell \&\ Wen 1992; [S~III], 
Wen \&\ O'Dell, 1993) at the Kitt Peak National Observatory Coud\'e Feed spectrograph at selected position angles
with respect to the Trapezium stars.  This work was used to compare the radial velocities with the
predictions of turbulence within a thin ionized layer (von H\"orner 1951) and the results are 
summarized in OD2001. A more ambitious and complete mapping of the radial velocities
is now being created as part of the Rice University PhD thesis of Takao Doi, who is creating velocity
images of the Huygens region at resolutions of 10 \kms\ and 2\arcsec\ in the \Ha, \OIII\ 5007 \AA, and \NII\
6583 \AA\ lines, and a similar study at the Instituto de Astronom\'{\i}a, UNAM, Morelia, M\'exico in the
red \SII\ doublet lines by W. J. Henney \&\ M. T. Garc\'{\i}a-D\'{\i}az that uses in part data obtained by
co-author O'Dell.

With the permission of our colleagues we have used their north-south orientation slits to create Figure 8
and Figure 9.  These are spectra taken from the velocity sample region indicated in Figure 1 and
used for the creation of the surface brightness profiles of Figure 7. The instrumental full width at half maximum intensity (FWHM) was very
close to 10 \kms\ in each case. Although the original samples along the slit were at intervals of
0.64\arcsec, the seeing was characteristically about 2\arcsec\ during the observations. The former
figure shows a half tone depiction of the spectra in lines characterizing various  parts of the
ionized layer and the latter presents contour diagrams of the same data. 

The power of this approach lies in the fact that various emission lines are characteristic of different
samples of the ionized layer. Although this fact is implicit in the discussion of several texts,
perhaps the best illustration is in a discussion of structure of the Helix Nebula (O'Dell 1998b).
\Ha\ emission comes from throughout the ionized layer, its emissivity being weighted towards the lower
\Te\ regions along the line of sight.  The forbidden line emissivity in \SII, \NII, and \OIII\ are
all weighted towards the higher \Te\ regions along the same line of sight.  Delineation of information
about structure along the line of sight is also provided by the ionization differences. \SII\ emission
will arise exactly in the ionization front, where ionization of hydrogen rapidly transitions from
nearly unity to nearly zero. A good fraction of the \NII\ emission will occur in the adjacent thin \hplus +\heneutral\ zone, while
the \OIII\ emission will arise in the outermost (closest to the observer) \hplus + \heplus\ zone.

We see in Figure 8 and Figure 9 that there is considerable velocity variation along the slit and that
the velocity in one ion is not well correlated with that in the other ions.  Thermal broadening of 
the \Ha\ line precludes a detailed analysis of its velocity structure, but is low enough in the
other, more massive,  ions to see the differences clearly.  One sees that the narrower the expected
emitting region, the more structure there is along the slit.  \SII\ emission gives evidence for
velocity structure with a scale of the seeing values of 2\arcsec\ and \OIII\ emission has a characteristic
size scale down to about 5\arcsec.  These data argue that structure exists within the main ionized
layer at sizes down to at least 2\arcsec\ and that gas of different states of ionization is moving
at different velocities. It is the spatially unresolved average velocity along the slits which 
give the characteristic progression to more blueshifted radial velocities at higher states of ionization that is the earmark of
a thin blister of gas expanding towards the observer.  

The second line of evidence for fine scale structure within the nebula is from comparison of
emission from very different states of ionization or conditions of emissivity. For this comparison we
used two pairs of images, whose closeness in wavelength essentially eliminate the effects of
interstellar extinction. The first pair 
used the F487N images, which are primarily due to \Hb\ emission whose emissivity is like that
of \Ha\ in that it selectively comes from the lower \Te\ regions, and the F502N images, which are
primarily emission from \OIII\ 5007 \AA\ high \Te\ collisionally excited regions.  The second pair
used the F656N \Ha\ filter and the \NII\ F658N filter. The considerations for the emissivity are similar
in the second pair of filters, but the emitting layer of the \NII\ emission should be thinner.
The hydrogen emission arises from throughout the ionized zone, but the \NII\ and \OIII\ emission arise
from selected zones, as described earlier in this section.

The results of these comparisons are shown pictorially in Figure 10.  One sees that there are 
wide variations in these ratios. Some of these changes are due to shocks caused by outflows from
pre-main sequence stars (O'Dell \&\ Doi 2003) while others such as the Bright Bar region in the 
southwest corner of the images is due to large scale topographic features of the 
ionization front (O'Dell \&\ Yusef-Zadeh 2000). In addition to these, there are many additional 
features.  A more quantitative depiction is shown in Figure 11 and Figure 12, where profiles along
the same region as the velocity samples are presented. We see fine scale structure in the ratios 
down to nearly the size of the 1\arcsec x2\arcsec\ bins over which the data was averaged, although a characteristic value would be 3\arcsec\ to 4\arcsec. The probable
errors of the ratios were derived from the total counts used in their calculation and are 1.8 \%\ for
the F658N/F656N ratio and 1.2 \%\ for the F502N/F487N ratio. As expected, the peaks are sharper in
the F658N/F656N ratio, indicating that the sources are small concentrated regions. The fact that there
is a general anti-correlation of the ratios indicates that the variations are largely due to 
ionization structure, rather than temperature differences.

In summary, we can say that both the spatially resolved high resolution velocity data and the line
ratio data indicate structure in the main ionized layer of 3\arcsec\ to 5\arcsec. 

A comparison with a profile of the results of the determination of \Tc\ is also useful and is shown
in Figure 13, where a profile along the same region as the velocity data is shown.  We see that the
proplyd 141-301 shows the common characteristic of the other proplyds, in having a higher than normal
formal value of \tfour, but, as shown in \S\ 4.3, this is probably due to collisional de-excitation of the 
electron state giving rise to the 5007 \AA\ emission. The ``JET'' feature shows a similar enhancement,
which may also be due to having densities comparable to the critical density for this nebular transition,
but it could be real, being the result of shock heating at the interface with the ambient nebular
gas. Aside from these features, there are variations in \Tc\ all along the slit.  In the well exposed
region from 50\arcsec\ to 150\arcsec\ we calculate that the probable error of \Tc\ is 0.5\%\, i.e.
45 K., which is small as compared with the observed variations.  There are no obvious general patterns
of changes with the extinction corrected surface brightness in \Ha, the velocity changes, nor the
line ratios.

\subsection{The effects of the ionization front advancing into a clumpy PDR}

There is considerable evidence that the PDR behind the ionization front of the Orion Nebula is 
very clumpy (Tauber, Goldsmith, \&\ Dickman 1991, Hogerheijde, Jansen, \&\ van Dishoeck 1995). 
HCN and HCO$^{+}$ observations by Young Owl \etal\ (2000) demand that the portion of
the PDR seen obliquely behind the Bright Bar feature possesses dense clumps (3 x 10$^{6}$ \cmq) 
embedded within an interclump medium of lower density (5 x 10$^{4}$ \cmq). More direct observations
come from imaging in \htwo. van der Werf \etal\ (1996) show that clumps exist from a scale of
40\arcsec\ down to their resolution limit of 1.5\arcsec. The highest angular resolution study in
\htwo\ is that of Vannier \etal\ (2001) who studied a region northwest of \thC\ at a resolution of
0.15\arcsec. Again, they saw clumpy structure, in this case going down to 1.5\arcsec -2.0\arcsec.
A clumpy structure in the PDR is not unexpected and there would be several instability mechanisms
that operate there (Capriotti 1973, Brand 1981, Vishniac 1994, Garc\'{\i}a-Segura \&\ Franco 1996,
Williams 1999). 

The implications of these clumps for the nature of the ionized layer of gas in the nebula are very
important since it is into these clumps and the interclump medium that the ionization front advances.
Where the density is lower, the advance will be more rapid. The expected rate of advance of the 
ionization front is not known as the hydrodynamic problem has not been solved.  The Orion Nebula
is old enough that it is in a state of quasi-equilibrium and the velocity will depend on many factors,
including whether or not the photoionizing star \thC\ has a velocity component towards or away from
the molecular cloud.

This means that one would expect the ionization front advancing into a clumpy PDR to move beyond the
dense clumps and be retarded at the clumps, so that a microscopic view of the ionization front would
be like that of a stubble-field. The region behind a dense knot would then be shadowed from 
ionizing Lyman Continuum (\LyC) radiation coming directly from \thC\ and would be illuminated only by diffuse (sometimes
called scattered) \LyC\ photons. Where the density is insufficient to form a quasi-static ionization
front, the clumps will simply retard the advance of the ionization front, while the dense clumps
will form a small local ionization front that will exist until it has lost sufficient material
through photoevaporation that the ionization front can penetrate, then totally disperse it.

\subsection{Low \Te\ regions behind clumps within the ionized gas of the ionized layer of gas}

The shadowed regions behind the knots within the ionized layer can be sources of low temperature
emission and of temperature variations, therefore they will increase the \tsq\ value. The photoionization 
equilibrium in such a shadowed 
region was considered 30 years ago (Van Blerkom \&\ Arny 1972, Kirkpatrick, 1972, Capriotti 1973), but it is 
only recently that the more general case has been considered (Cant\'o \etal\ 1998, henceforth C98).
The region illuminated only by diffuse LyC will see a lower temperature radiation field because most
of the recombinations of \hplus\ that produce the diffuse LyC will result in a photon just above the
ionization threshold, while the direct stellar field will have more high energy photons and produce
higher \Te\ regions. C98 have solved the general problem, rather than trying to apply their calculations
to only a specific object, including the role of ambient gas density and the time variability of the
shadowed zone structure. When the combination of the stellar radiation density (the diffuse radiation
field density is about 15 \%\ of that) and gas density is right an ionization front
can form immediately at the edge of the shadowed cylinder of gas.  As the radiation density increases
and/or the density decreases, the ionization front can form within the shadowed cylinder and even
produce a condition where the shadowed cylinder is completely photoionized. C98 argue that \Te\ within
the ionized shadow zone will be about 6000 K, considerably cooler than that of the directly illuminated gas.

The radiation from the shadowed zone can be seen by the distant observer for two reasons. In some 
cases the angle between the line of sight from the observer to \thC\ can be quite different from the
angle between \thC\ and a local spot on the ionization front, allowing direct view of the shadowed
region.  Even when that is not the case, a significant fraction of the radiation from the shadowed
region will be scattered by dust in the nearby PDR, in the same way that radiation from the ionized
layer is scattered and produces the observed redward asymmetry of the emission lines (O'Dell, Walter, \&\ Dufour 1992 , Henney 1998).  

\subsubsection{Proplyds}

Shadowed regions have already been detected in the Orion Nebula. In a study of both the Cometary Knots in the Helix 
Nebula and 
the proplyds in the Orion Nebula, O'Dell (2000) showed that the proplyds cast ionization shadows 
as long as 0.2 pc.
The proplyds are optically thick to LyC radiation (O'Dell 1998a) and in that sense resemble the optically
thick clumps we argue exist at the main ionization front of the nebula. The proplyd ionization shadows
are most visible when the geometry is such that the shadow falls into the ionized layer that provides
most of the radiation identified as the nebula. O'Dell (1998a) showed that the shadows of the proplyds
covered about 0.5\%\ of the view from \thC . By assuming: pressure equilibrium, $1- \gamma=0.4$, 
$T$(N$^+$) = 10,100 K (B91, EPTE98, R2003) for the directly 
illuminated regions, and $T$(N$^+$) = 6,000 K (C98) for the shadowed regions we obtain 
a contribution of 0.0093 to \tsq (O$^+$) (see equation (10)), that is a contribution of 0.0043 
to the total \tsq\ (due to second term of equation (19)) . To reach the estimated 0.011 contribution to 
\tsq\ due to the O$^+$ zone, we need an additional 0.7\% of the volume of the nebula to be in a different type 
of ionized shadowed region.

\subsubsection{Neutral high density clumps}

In what follows we will propose the presence of high density clumps inside
the ionized layer left behind by the advancing ionization front. 
In the model we propose (illustrated in cartoon form in Figure 14) the clumps would not be permanent,
rather, they would exist in the ionized layer only until they are destroyed by photoevaporation. 
The cool columns of gas behind them would then come to the higher equilibrium \Te\ appropriate for
illumination by \thC\ rather than diffuse LyC.

One can approximate the results of an ionization front advancing into a clumpy
PDR, as described in \S\ 7.4.  We don`t know the velocity of this advance. It must be small,
because the tracers of the ionization front ([O~II] and [S~II]) are blueshifted 2.5 \kms\ with 
respect to the PDR tracers (CO, C~II) (O'Dell 2001a).  This blueshift is probably due to the
hot gas tracers coming from a layer that has already been accelerated by the pressure
gradient, rather than being the result of \thC\ moving away from the PDR. The radial velocity of \thC\ 
is highly uncertain because of variability probably arising in the extended atmosphere of the 
star. It is unlikely that this star is moving at a high velocity with respect to the other
cluster stars and their velocity is the same as the PDR (O'Dell 2001a) within the uncertainty of 3 \kms.  
In the discussion below we will assume that the ionization front is advancing into the PDR at
a rate of 2 \kms, but it should be recognized that this could be off by a factor of two.
In \S\ 7.5 we showed that the fluctuations in the surface brightness of the nebula indicate structure
with a height of about 0.025 pc. This would mean that the characteristic clump will have been 
illuminated by LyC radiation for about 1.3x10$^{4}$ years, which we will call the kinematic age.

If we can determine the column density and rate of loss of atoms per unit time and area,  we
can calculate the time that the clumps can survive photoevaporation and have a check on our basic model.
If the clump diameters are about 0.0044 pc (corresponding
to 2\arcsec) and the densities are 3x10$^{6}$ \cmq\ (Young Owl \etal\ 2000) the average column density of
a clump would be 2.8x10$^{22}$ atoms \cmsq.
The corresponding mass would then be about 7x10$^{30}$ g).

The photoevaporation rate for these clumps will be very similar to those of the proplyds, which
Henney \etal\ (2002) have shown to be 8x10$^{-7}$ \subsun\ \yr. Since these objects are comparable 
in size to the clumps at the ionization front, their average rate of loss of atoms is relevant, this 
being 2x10$^{11}$ atoms \cmsq\ \s. In this case, the lifetime of the clumps would be about 5,000 yrs.
However, the proplyd measured by Henney \etal\ (2002) must be much closer to \thC\ than the ionization
front, so that the rate of loss has been overestimated and the survival age of the clumps underestimated.
This means that the survival age of the clumps and their kinematic ages are compatible. They are, however,
close enough that the process involved must be a dynamic one, with any one clump surviving for only
about 10$^{4}$ years, but the number of clumps present at any time would be about constant as new
clumps are revealed by the advancing ionization front.

We can estimate the effect of these shadowed regions if we know their emissivities and volumes.
If the shadowed regions exist long enough that dynamical equilibrium is established with their 
surroundings (C98), then their volume emissivity will be four times that of the ambient lower density and
higher \Te\ gas.  If pressure equilibrium is not established, then the emissivity increase will be
about two times that of the ambient gas. If one interprets the non-JET fluctuations in the extinction
corrected \Ha\ surface brightness (Figure 7) as being due to knots at various distances from \thC, then 
they lie in a range of about 0.025 pc or 11\arcsec, which is comparable to the emissivity scale-height
of 15\arcsec\ derived in \S\ 7.2. This means that the shadowed regions occur within the zone where most
of the emission occurs.

To calculate the contribution to \tsq (O$^+$) and \tsq\ due to the shadows behind these 
clumps we need an estimate of the fraction of the emission produced in the shadowed regions. 
Assuming: a) the typical clump density to be $N_e = 3 \times 10^6$ \cmq\ , b) the average density of the ionized
media to be $N_e = 5 \times 10^3$ \cmq\ , and c) that half of the ionized matter originated from the photoevaporation
of these clumps, it follows that the volume originally occupied by the clumps must be about 
1/1200 of the total volume. However, the shadowed volume is larger than the volume occupied by the clumps;
we know that the shadows are approximately 0.025 pc long, while each clump is approximately 0.0044 pc across, 
so the shadowed volume can amount to approximately 1/200 of the volume of the region 
closest to the main ionization front, before these clumps are photoevaporated. 
Since these clumps survive long enough to cast 0.025 pc shadows, and the emissivity scale-height 
of the Orion Nebula is of 0.034 pc the total volume of the shadowed regions is about 1/250 of the  
effective emitting volume.

There are three possible physical conditions in the shadowed
regions: a) that they are completely neutral, in this case they
will not contribute to \tsq (O$^+$) nor to \tsq ; b) that they are
completely ionized, but not in pressure equilibrium with their
surroundings, then their contribution to \tsq (O$^+$) would be about
0.0038 and to \tsq\ of about 0.0016; and c) that they are completely
ionized and in pressure equilibrium with their surroundings, then
their contribution to \tsq (O$^+$) would be about 0.0075 and to \tsq\
of about 0.0032.  The contribution of the shadows, including the shadows of
the clumps and of the proplyds, amounts to $0.009 \la t^2 ({\rm O}^+) \la
0.017$, that corresponds to $0.0043 \la t^2 \la 0.0075$. Note that in
the shadowed regions oxygen is either neutral or once ionized, and
that therefore the shadowed regions do not contribute to the \tsq
(O$^{++}$) value. 
 
It can be shown that the ionized part of the proposed clumps has
a very small effect on \tsq. The clumps are expected to be close to the PDR,
and  their ionized part will have a larger density than the
interclump medium, so we expect the oxygen in the ionized fraction of the
clumps to be singly ionized.  Additionally the clumps cover a total solid
angle of about 0.03 steroradians, and therefore absorb less than 1/400 of
the ionizing photons, about 1/800. This implies that their emissivity
amounts to about 1/800 of the total emissivity of the nebula and that
the physical conditions in the high density photoevaporated region
will produce a negligible effect in the increase of \tsq (O$^+$) or
\tsq.

\subsection{Can the clumpy model for the Ionized Layer explain the Anomalous Line Broadening in Orion?}

There is an unexplained component of line broadening in the Orion Nebula that may be explained by
the clumpy model for the ionized layer that we propose. This anomaly first surfaced with Casta\~neda's
(1988) study of \OIII, and remained with the inclusion of other massive ions (\OII, Jones 1992;
\OI, O'Dell \&\ Wen 1992; \SIII, Wen \&\ O'Dell 1993). After 
quadratic subtraction of the instrumental, thermal, and turbulent components of  broadening, there 
remains a large and unexplained line width having FWHM$\simeq$10 \kms. The large thermal width
of the \HI\ lines makes it difficult to accurately determine the extra width there, however, Wilson,
\etal 's (1997) study of the H64$\alpha$ line indicates an unexplained broadening component (called the
turbulent velocity in their paper) of 19.6$\pm$0.9 \kms.

This extra broadening component has been re-addressed  using spectra
made with the Keck telescope for the study of proplyds and their jets (Henney \&\ O'Dell 1999, Bally, O'Dell, \&\ McCaughrean 2000. 
The Keck data were very useful as they have the highest velocity resolution (6 \kms)
of any of the optical wavelength studies and allow
examination of all of the primary lines of \OI, \OIII, \SII, \NII, He~I, \Ha, and \Hb\ at exactly the
same positions. Samples were taken from 14\arcsec\ long spectra centered on 137-349, 150-353, 163-357, 170-337,
177-341, 182-413, and 244-440 (the designation system is described in O'Dell \&\ Wen 1994) and 
28\arcsec\ long spectra taken at the positions 137-349, 150-353, and 163-357.
After correction for instrumental,
thermal, and turbulent (assumed to be 1  \kms) broadening, the remaining line widths (FWHM in \kms) were
\Ha\ 20.9$\pm$1.3, \Hb\ 18.8$\pm$2.2, He~I 5876 \AA\ 18.4$\pm$2.9, \NII\ 6583 \AA\ 10.6$\pm$1.4, \OI\ 6300 \AA\ 9.0$\pm$2.1,
\OIII\ 5007 \AA\ 13.0$\pm$3.7, \OIII\ 4959 \AA\ 11.3$\pm$2.1, \SII\ 6731  \AA\ 11.3$\pm$2.4, and
\SIII\ 6312 \AA\ 11.8$\pm$1.9. The only source of [O~II] data are those found in the thesis of 
Michael Jones (1992) who found for 10.5$\pm$2.5 for the central 80\arcsec x 80\arcsec.
The longer Keck slit spectra were subjected to a separate analysis using 
various fractions of the total slit length and no variation of the results were seen, i.e. there was
not a relation between sample size and the residual line width. The radio (H64$\alpha$, 19.6$\pm$0.9) and 
optical (\Ha, 20.9$\pm$1.3 and \Hb, 18.8$\pm$2.2) are in agreement within their individual uncertainties.

The unexplained broadening in the emission lines that arise from the lower temperature component
of the nebula is twice that of lines arising from the higher temperature component. We see that the
recombination lines of H~I and He~I that preferentially come from cool regions have a clearly larger unexplained width (average 19.4 \kms)
than the collisionally excited forbidden lines that preferentially come from hot regions (average 10.9 \kms).
This strongly argues against the broadening arising from Alv\'en waves (OD2001) since they would have little
dependence upon \Te. The explanation certainly does not lie with there being a systematic difference in
\Te, because the sense of the correlation is opposite to that expected.

The clearcut differences in the extra broadening in collisional and recombination lines is probably
related to the the mechanism giving rise to the temperature inhomogeneities. Can the broadening be due to the clumpy structure that we argue 
can explain the temperature inhomogeneities?  O'Dell \&\ Wen (1992) argue that clumps having sizes
of less than 2.3\arcsec\ could produce the extra broadening, but this argument was made before 
knowledge of the larger broadening of the hydrogen lines. The steady process of destruction of the
clumps by photoevaporation would certainly give rise to flows away from the clumps. However, the line
broadening due to this should selectively be a blueward distortion of the integrated line-of-sight line profile,
 except for the final stage of destruction, when material would be flowing in all directions.
However, the initial phase of a shadowed region would be accompanied by material flowing
inward towards the low pressure zone that would exist when the shadowed \Te\ drops and the shadowed region is moving towards
a state of hydrostatic equilibrium. It is obvious that these situations need to be theoretically 
modeled. Although it is not now possible to quantitatively link them, it is likely that the explanation of the electron temperature fluctuations and the line
widths are related.

\section{Conclusions}

{From} our $T_c$(O$^{++}$) map of the Orion Nebula, that includes
1.5 x 10$^{6}$ independent temperature determinations, we have found that
\tsqa (O$^{++}$)$= 0.008$.

{From} our \tsqa (O$^{++}$) value, together with geometrical considerations and other
observations in the literature, we estimate that \tsq (O$^{++}$)$=
0.021$.  Note that the total  \tsq (O$^{++}$) is larger
than \tsqa (O$^{++}$) because in addition to the variations across the
plane of the sky it includes the temperature
variations along the line of sight.

{From} our \tsq (O$^{++}$) value and comparisons between the
temperatures in the low- and high-ionization zones, the O$^{+}$ and O$^{++}$
zones, we find that \tsq (H~{\sc ii})$ = 0.028 \pm 0.006$. 
\tsq (H~{\sc ii}) is different to \tsq (O$^{++}$) because in 
addition to the variations in the O$^{++}$ zone includes the variations in the
O$^{+}$ zone and the difference in the average temperature between both
zones. Our derived \tsq (H~{\sc ii}) value is 7 times higher than those
obtained from homogeneous one-dimensional photoionization models of
the Orion Nebula.

A combination of sources needs to be found to
explain the large $t^2$ values observed in the Orion Nebula.
Many possible causes of temperature variations have been discussed in the
literature like: density variations, deposition of mechanical energy,
deposition of magnetic energy, presence of shadowed regions,
chemical inhomogeneities, dust heating, and transient effects due
to changes in the ionizing flux. 

{From} observations of variations in surface brightness, columnar electron temperature,
radial velocity, and ionization, we have established the presence of
significant small scale structure at the 1-5 arc sec level. The small
scale velocity and ionization structures are probably strongly linked
with the temperature structure.

The presence of this small scale structure is compatible with a
new model for the structure of the nebula near the ionization
front. We posit that this front is advancing into the highly clumpy
photodissociation region revealed by radio and infrared
observations. These clumps possess ionized shadows that are of low
\Te . A quasi-static equilibrium exists in which any one clump
survives the process of photoevaporation only for a short time.  The
process may be common to most \HII\ regions which have a portion that
is ionization bounded. There is an unexplained component of line
broadening in the Orion Nebula that may be explained by the clumpy
model for the ionized layer that we propose. 

We have discussed the presence of shadowed regions due to proplyds and
to high density clumps near the PDR. We estimate that the shadowed
regions can contribute $0.0043 \la t^2 \la 0.0075$ to the
total $t^2$ value, which together with the homogeneous one dimensional
ionization structure provide only from about one third to about half
of the observed $t^2$ value. Additional sources of temperature
variations are needed to explain the total $t^2$ observed value. 
Moreover, since in the shadowed regions O is only once ionized, these
regions could be responsible for temperature variations in the O$^+$
zone, but not in the O$^{++}$ zone, and since the \tsq (O$^{++}$)
value is very large we need one or several additional sources of
temperature variations to explain it. 

There is an anomalous broadening mechanism operating in the Orion Nebula
which is unexplained. It broadens H and He recombination line emission by about 19 \kms\
and the forbidden line emission from heavy ions by about 11 \kms. Since these
two types of lines selectively rise from the low and high temperature components 
along a line of sight, it appears likely that this process is related to 
the \tsq\ problem.

\acknowledgements

Michael Richer and Will
Henney provided valuable support in obtaining the high velocity
resolution spectra at the Mexican National Observatory at San Pedro
Martir.  Will Henney also provided several useful comments on an earlier 
version of this paper.
Takao Doi kindly shared before publication the results of his
Orion Nebula radial velocity program, which allowed us to generate
Figure 8 and Figure 9. Henney also provided observations on the expected
rate of motion of the ionization front in the Orion Nebula.
Gary J. Ferland provided estimates for the heating rates. 
Christopher A. Coco measured the Keck spectra
line widths as part of his senior thesis at Rice University.  The
authors are grateful to Patrick Hartigan of Rice University and Robert
Rubin of the Ames Research Center for discussions on some aspects of
this work. CRO's work was supported in part by grant GO-9141 from the
STScI to Vanderbilt University.  MP's work was supported in part by
grant IN 114601 from DGAPA UNAM.

\clearpage

\begin{deluxetable}{lccccccl}
\tabletypesize{\footnotesize}
\tablecolumns{8}
\tablenum{1}
\tablewidth{0pt}
\tablecaption{Filter and WFPC2 Characteristics}
\tablehead{
\colhead{Filter}  & \colhead{F437N}  & \colhead{F469N}  & \colhead{F487N}  & \colhead{F502N} & \colhead{F547M}  & \colhead{F656N} & \colhead{F658N}} 
\startdata
E$\rm _{filter}$\tablenotemark{a} & 16.56 & 17.40 & 19.87 & 22.79 & 601.9 & 22.0
4 & 31.19 \\
$\rm T^{filter}_{l1}$~\tablenotemark{b}     &  0.50 &  0.52 &  0.59 &  0.64 &   
--  &  0.78 &  0.79 \\
$\rm T^{filter}_{sec}$~\tablenotemark{c}    & 0.005 &   --- &   --- &  ---  &   
--- &  0.24\tablenotemark{d} &  0.045 \\
s~\tablenotemark{e}&0.061  & 0.074 & 0.084 & 0.091 & 0.125 & 0.145 & 0.145  \\
\tablenotetext{a}{
$\rm E_{filter}=\int_{0}^{\infty} T_{\infty} \mbox{d}\lambda$=effective
width of the filter in \AA.}
\tablenotetext{b}{
$\rm T^{filter}_{l1}$=transmission at primary line for the filter.}
\tablenotetext{c}{
$\rm T^{filter}_{sec}$=transmission at the secondary line for the filter.}
\tablenotetext{d}{
This value is for the [N~II] 6548 \AA\ line, it is 0.055 at the three times stro
nger [N~II] 6583 \AA\ line.}
\tablenotetext{e}{
The WFPC2+HST system throughput determined in 2001 June from Table 2.4 of the WF
PC2 Instrument Handbook.}
\enddata 
\end{deluxetable}

\clearpage

\begin{deluxetable}{lrrr}
\tablecaption{$t^2_A({\rm O}^{++})$.}
\tablewidth{0pt}
\tablehead{
\colhead{Region} &
\colhead{Number of pixels $(10^6)$} &
\colhead{$\left< T_c \right>$} &
\colhead{$t^2_A({\rm O}^{++})$} 
}
\startdata
Region 1 & 0.51 & 9190 & 0.0117 \\
Region 2 & 0.43 & 9390 & 0.0156 \\
Region 3 & 0.34 & 9240 & 0.0050 \\
Region 4 & 1.50 & 9250 & 0.0096 \\
Region 5 & 1.09 & 9220 & 0.0084 \\
\enddata

\end{deluxetable}

\clearpage

\figcaption[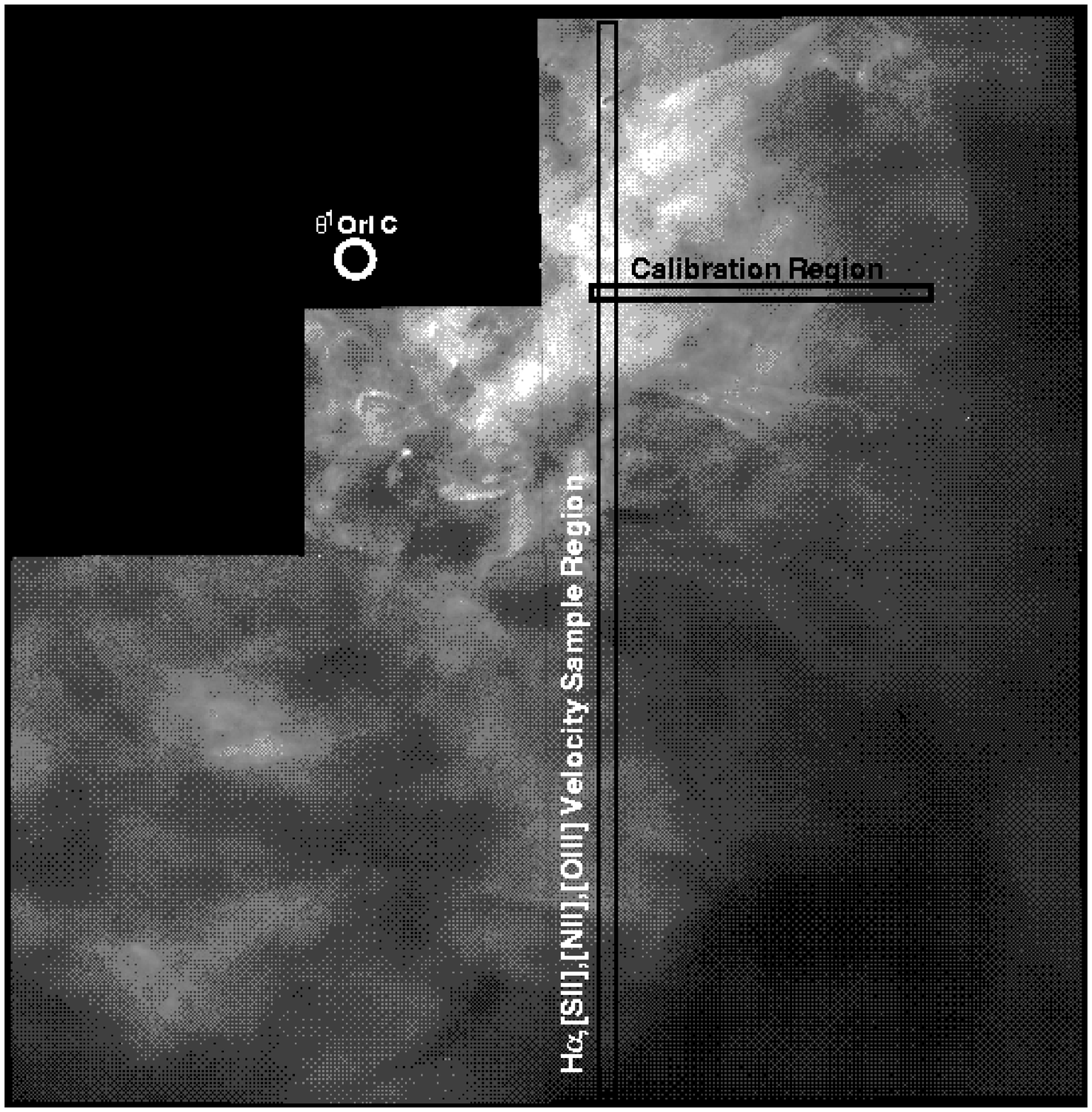]{This \OIII\ 5007~\AA\ image from the data set illustrates
the location of the WFPC2 field of view. The horizontal width of the image is
149\arcsec\ and the central intersection of the edges of the CCD detectors is
at 5:35:14.2 -5:24:03 (2000). The vertical axis points north. The regions used
for calibration and for comparison with spectroscopic and ionization changes
are indicated.}

\figcaption[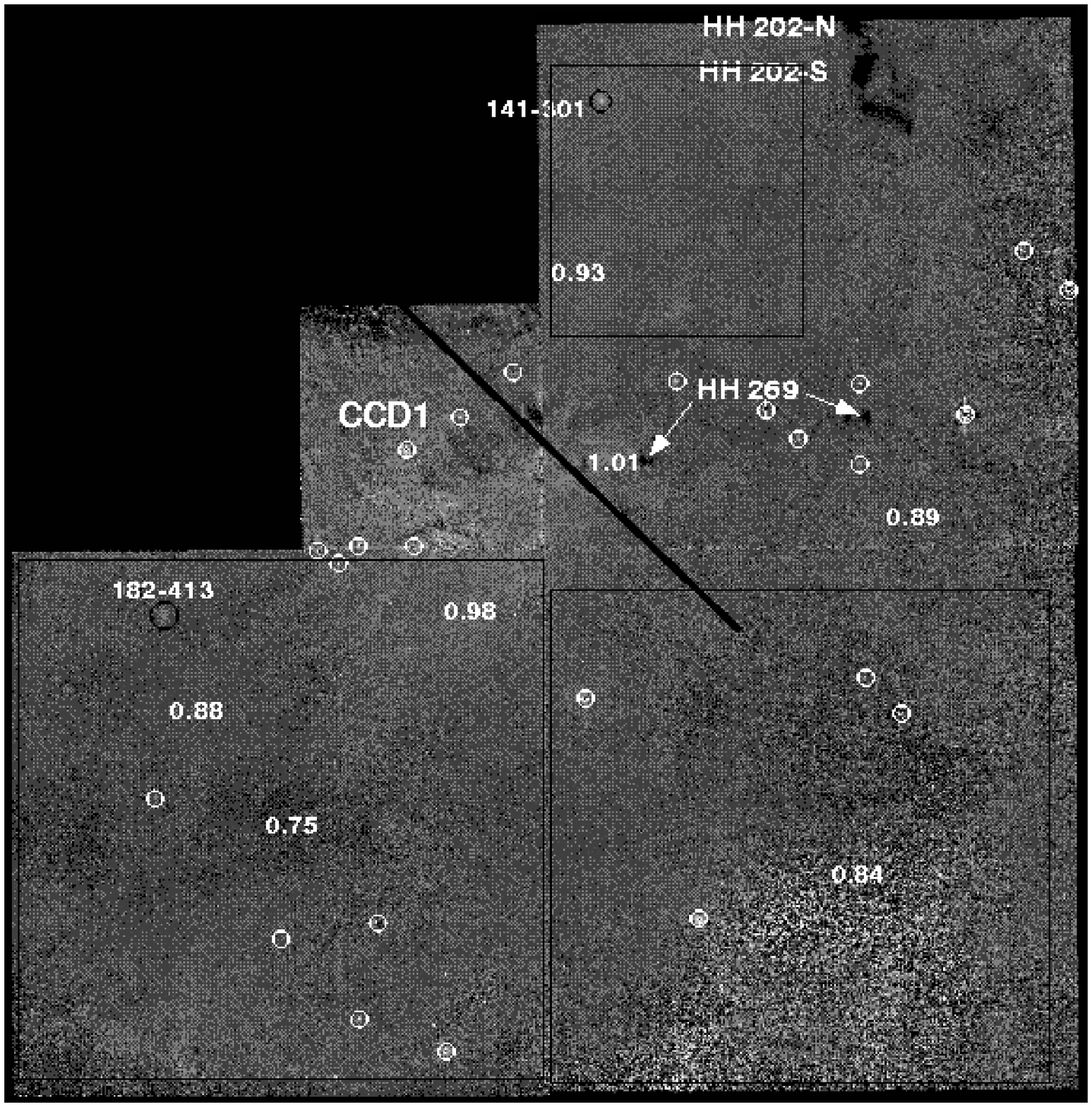]{This is the \Tc\ image derived from the WFPC2 dataset for
the field of view shown in Figure 1, at the same scale and orientation. Local
values of \Tc /10,000 are shown as numbers like 0.93. The white open circles
surround features that are ``scars'' caused by stars. Two proplyds that show
high values of \Tc\ are labeled (141-301, 182-413).  Features associated with
HH~202 and HH~269 are labeled and discussed in the text. The dark solid line
lies over a diffraction spike caused by one of the vanes holding the secondary
mirror of the HST. Features within CCD1 are suspect because of scattered light
from \thC\ and other bright Trapezium stars.  The black outlines depict the
regions assessed for small scale changes in \Tc\ as shown in Figure 3.}

\figcaption[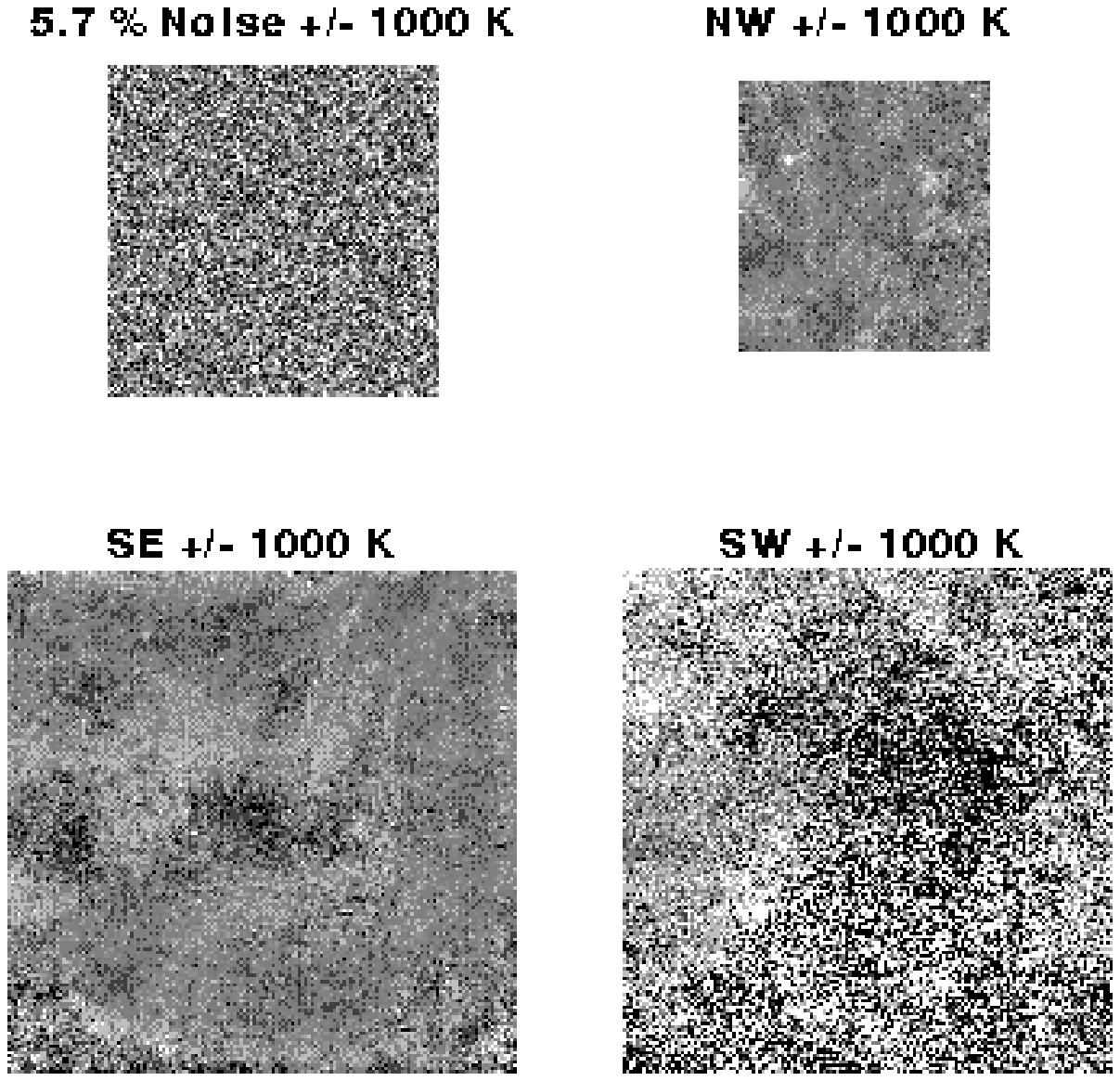]{The regions outlined in Figure 2 are shown here in the same
spatial orientation. A Poisson-noise-only image of 5.7 \%\ noise is shown in
the upper left.  The display range has been adjusted to show the intermediate
scale fluctuations in \Tc.  The pure noise field is 50\arcsec x 50\arcsec.  The
northwest (NW) field is 37\arcsec x 40\arcsec and the southeast (SE) field is
76\arcsec x 76\arcsec.  The southwest (SW) field is 72\arcsec x 75\arcsec and
its resolution is one half that of the other images, which are one pixel of
0.0996\arcsec.  Because the southeast region is about one fourth the surface
brightness of the other fields, its data were averaged in bins of 2 x 2 pixels,
thus making the expected noise comparable to that of the other images.}

\figcaption[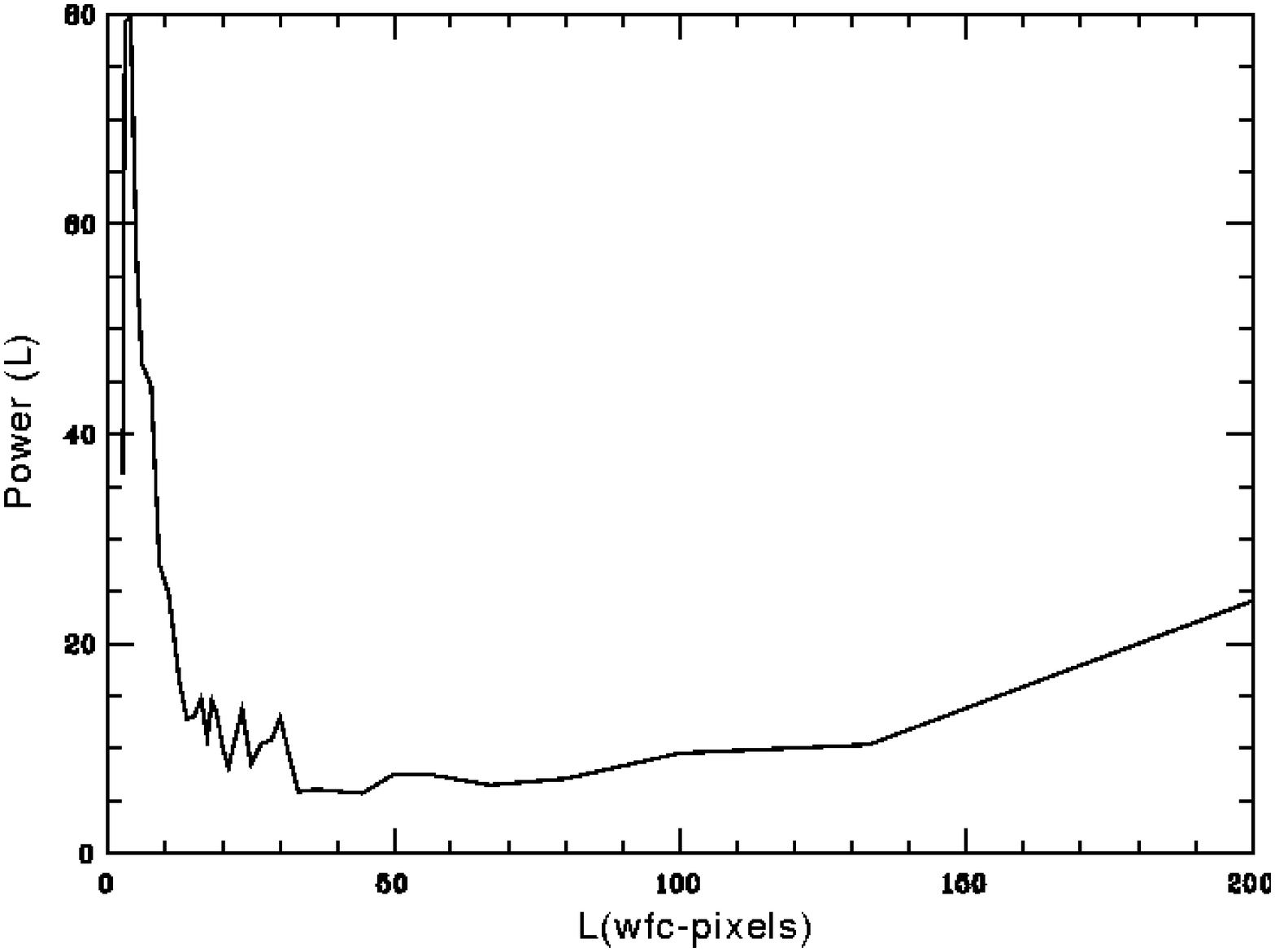]{The results of a Fourier analysis of a section of the \Tc\
image in CCD2 is shown as a power spectrum with noise subtracted in arbitrary
units (ordinate) versus the size scale (L) in units of the coarser WFPC2 pixels
of 0.0996\arcsec.  One sees the highest power in the smallest scales, although
there is power even in the largest scale.}

\figcaption[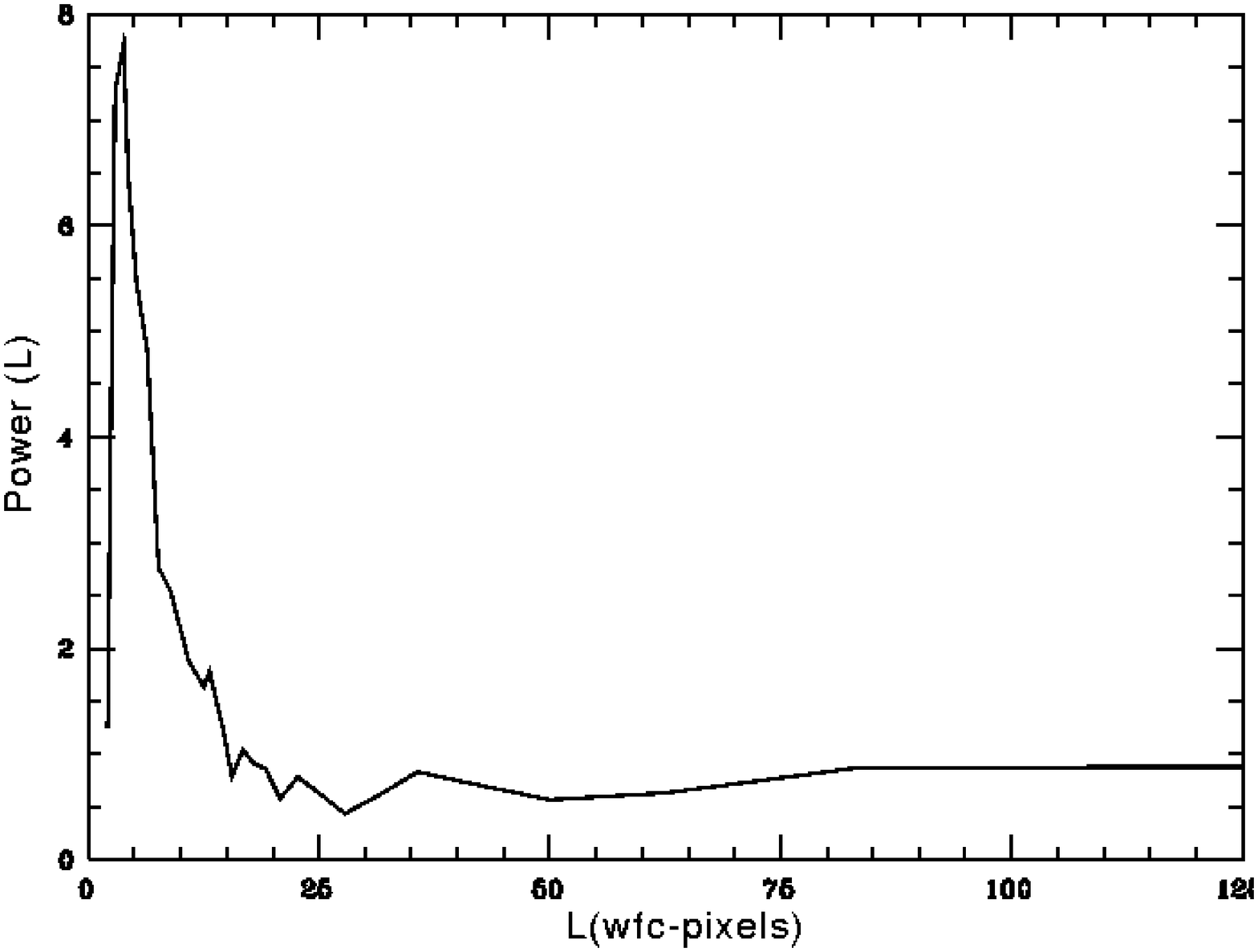]{The same as Figure 4, except for a region selected from
CCD4.}

\figcaption[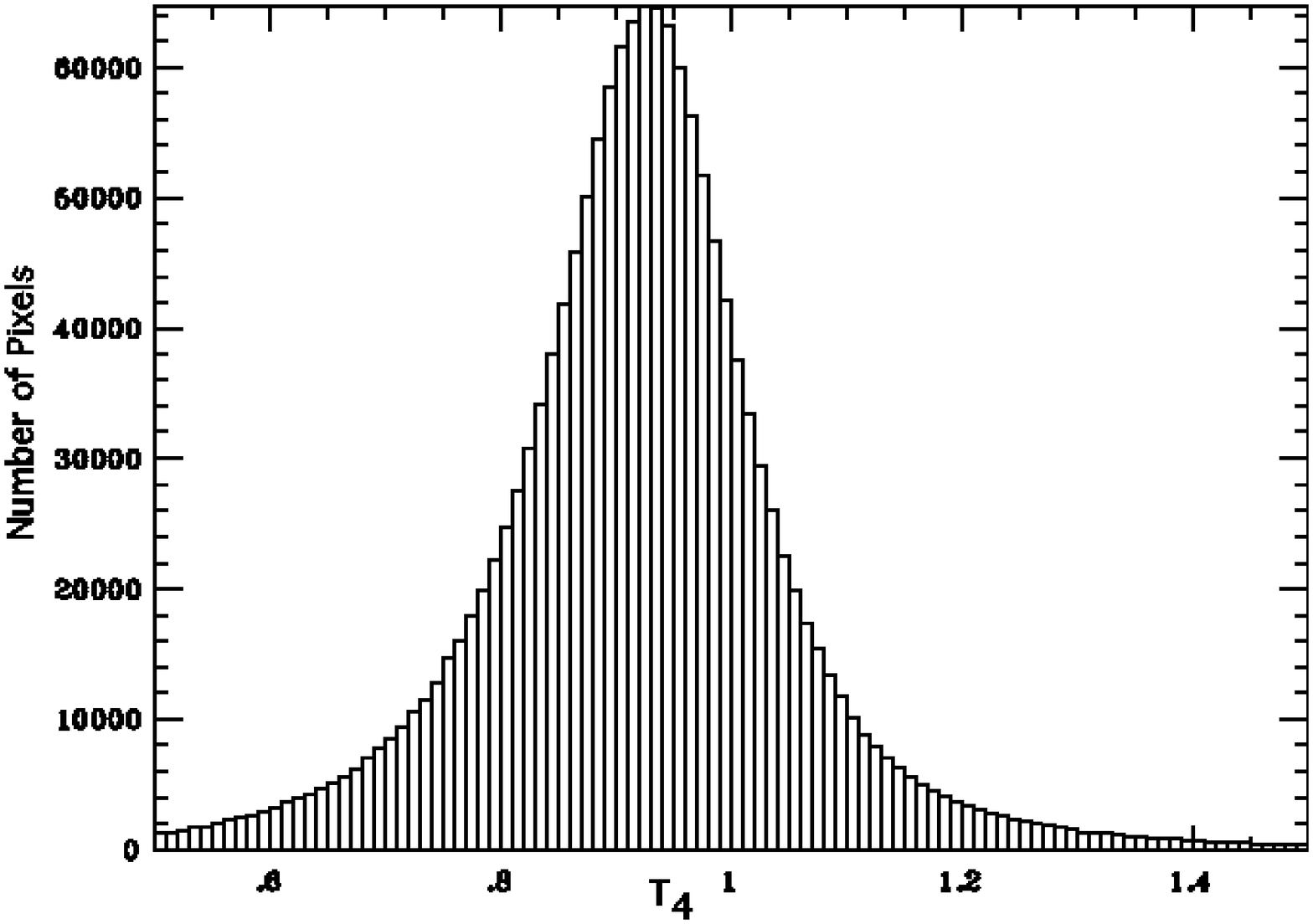]{Histogram of the $T_4$(O$^{++}$) columnar values from the
pixels in region 5. Note that \tsqa\ (O$^{++}$) cannot be obtained directly
from this histogram as each pixel needs to be weighted by its emmision measure,
see text.}

\figcaption[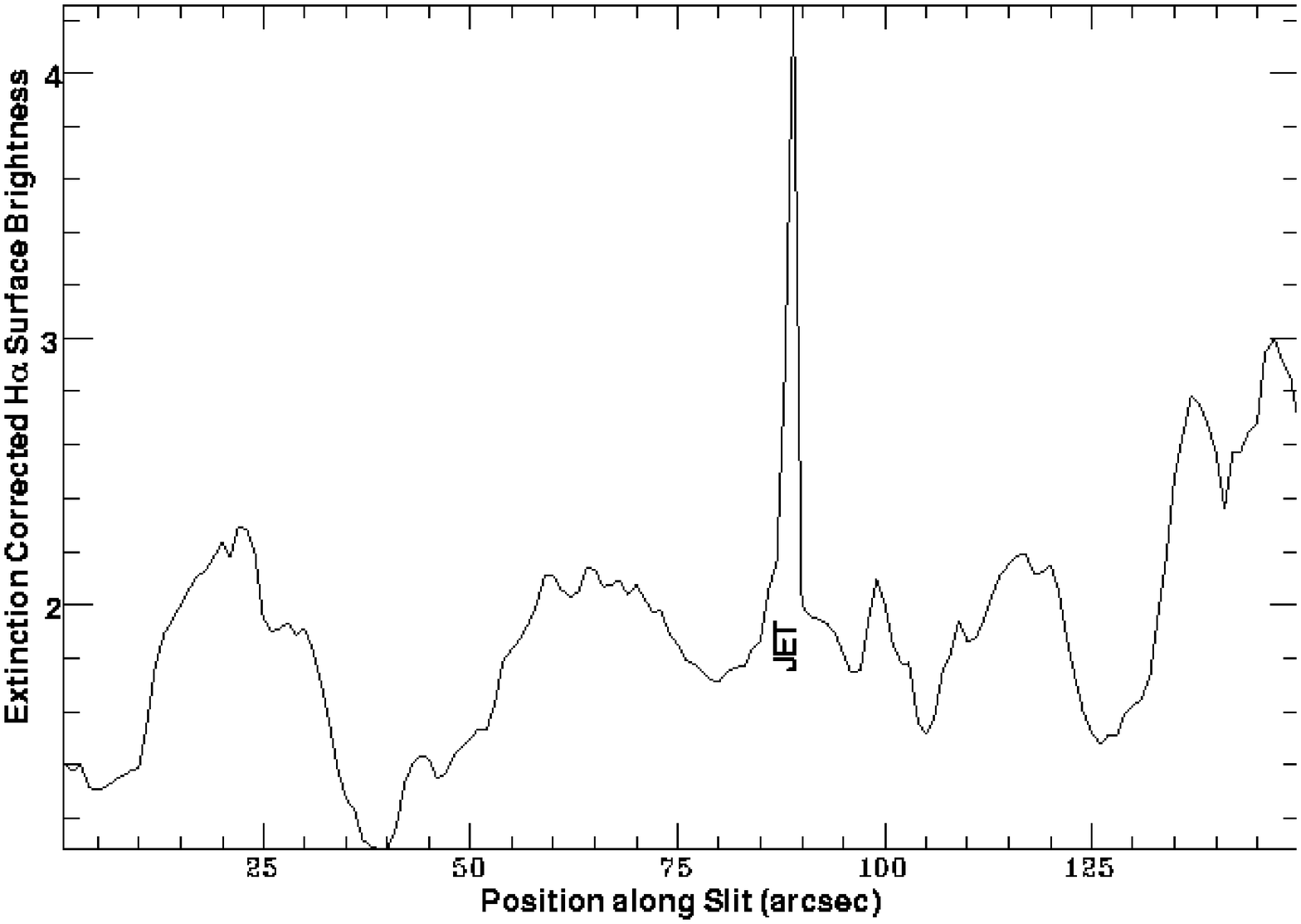]{The surface brightness in the \Ha\ line, after correction
for extinction by the radio-optical method of O'Dell \&\ Yusef-Zadeh (2000), is
shown as a function of distance (increasing numbers to the north) for the
velocity sample region shown in Figure 1. The featured marked ``JET'' is the
inner jet associated with HH~269 (O'Dell \&\ Doi 2003).}

\clearpage

\figcaption[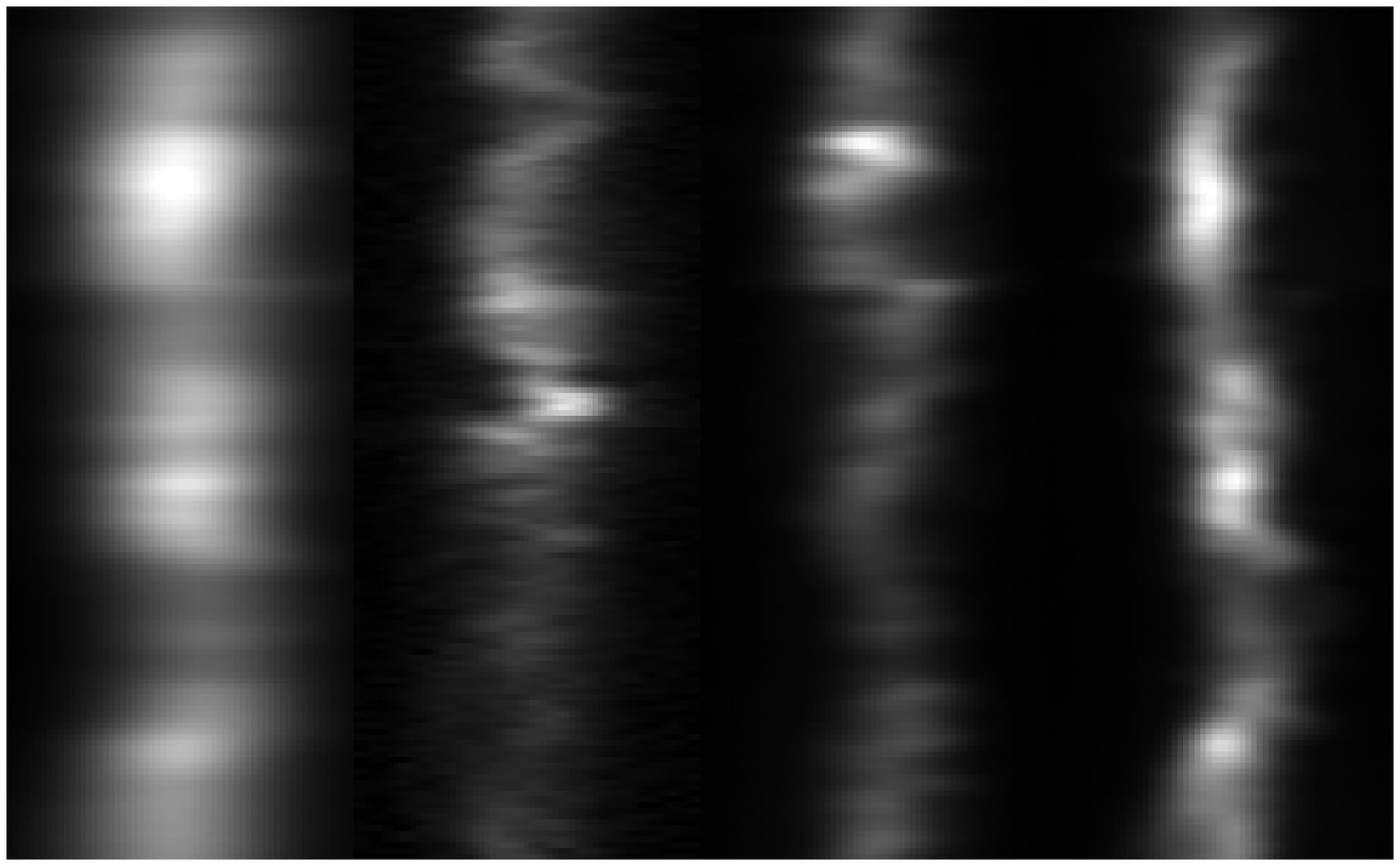]{The emission from four lines (from left, \Ha, \SII, \NII,
\OIII) are shown for the sample indicated in Figure 1, with north at the
top. The additional thermal broadening of the \Ha\ line causes its spectrum to
be noticeably wider than the more massive ions. One sees a progression of fine
scale motion as the volume of the emitting zone is decreased, \Ha, \OIII, \NII,
and \SII.}

\figcaption[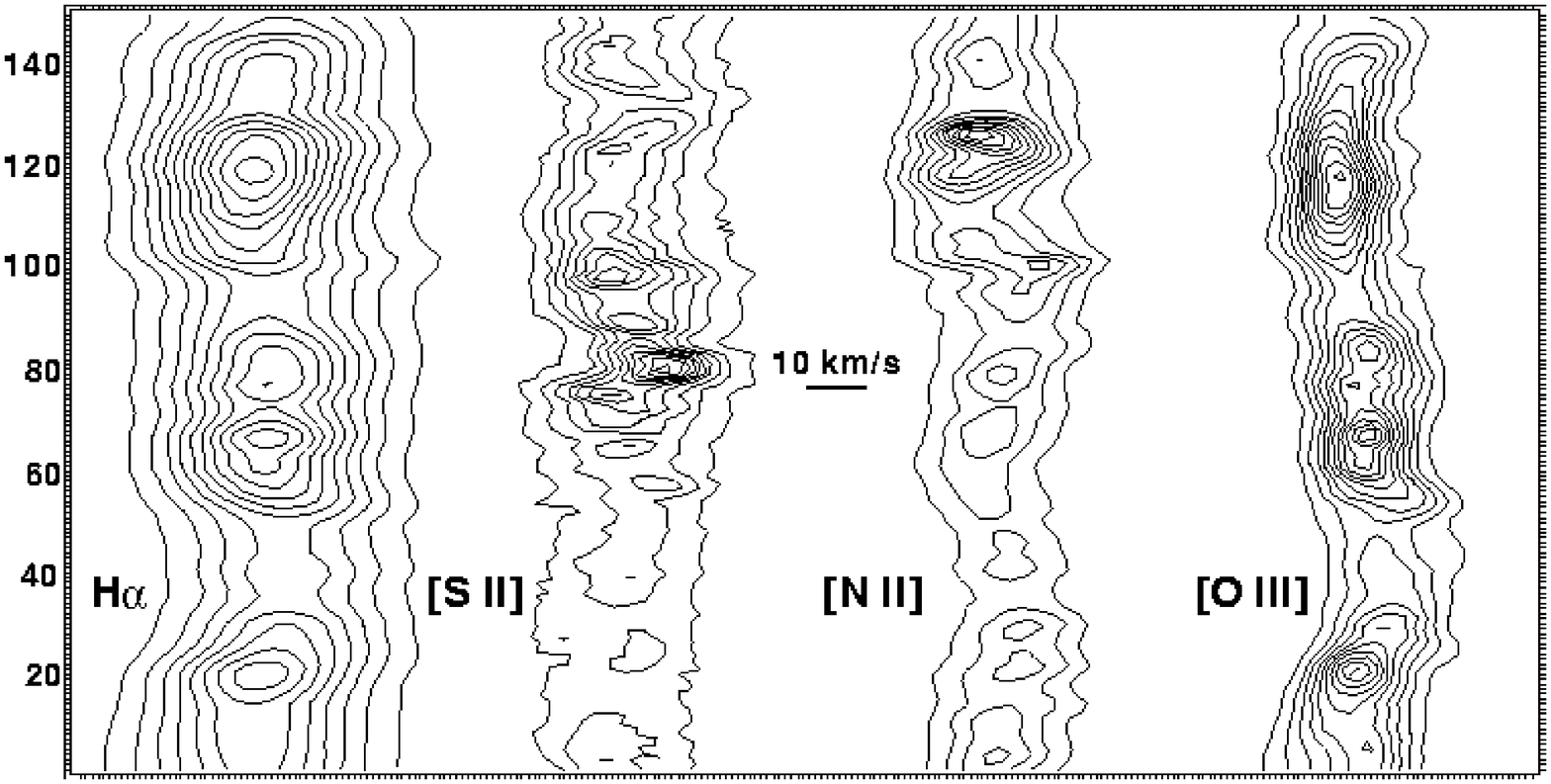]{This linear, equal interval, contour plot shows the same
spectral data as in Figure 8.}

\figcaption[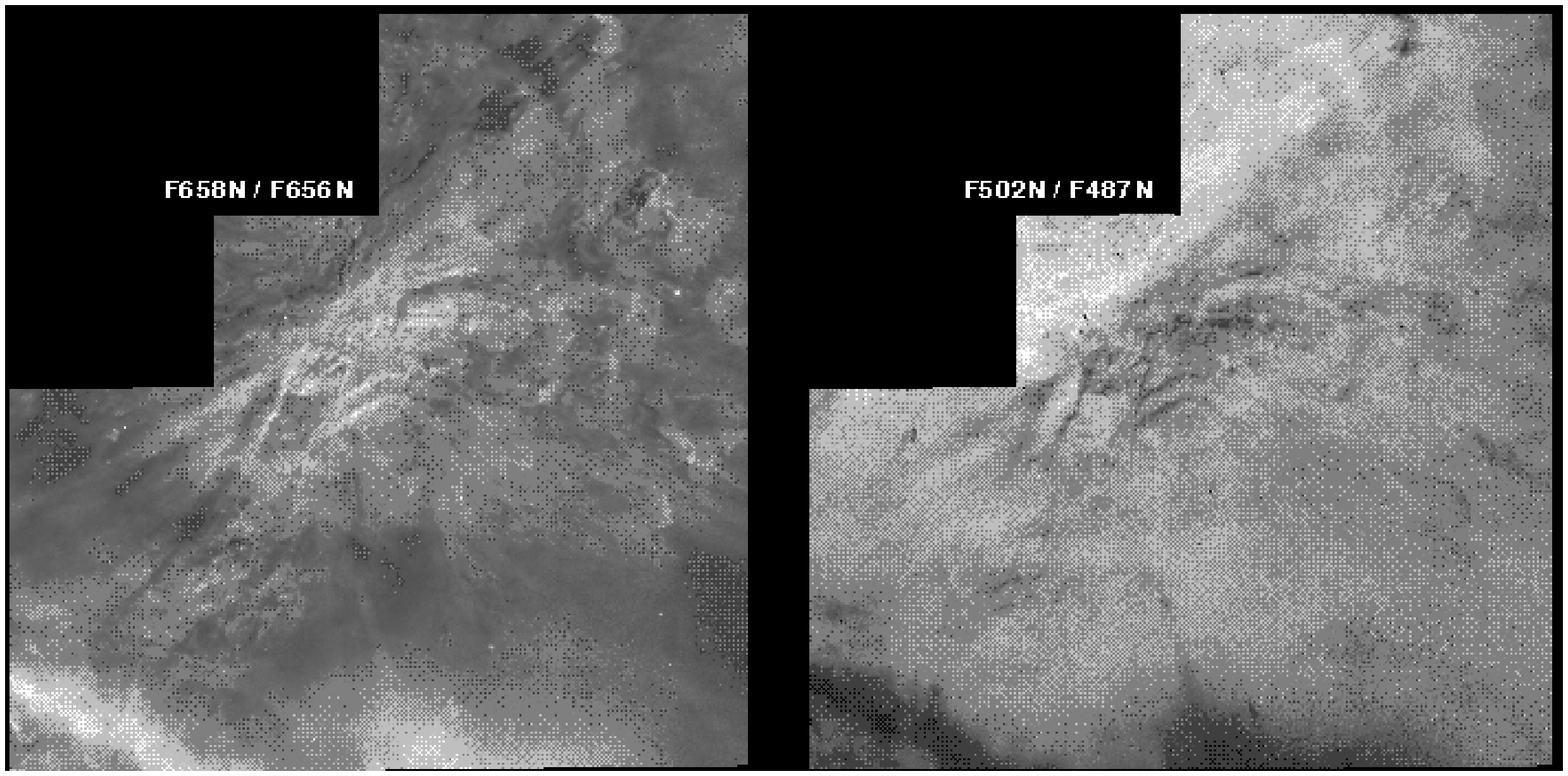]{These figures show the same field of view as Figure 1. The
left hand side shows the ratio of the signal in the F658N filter divided by the
signal in the F656N filter. The right hand side shows the ratio of the signal
in the F502N filter divided by the signal in the F587N filter. Both images are
intended to show small scale ionization changes and to do this in a way
insensitive to varying amounts of interstellar extinction.}

\figcaption[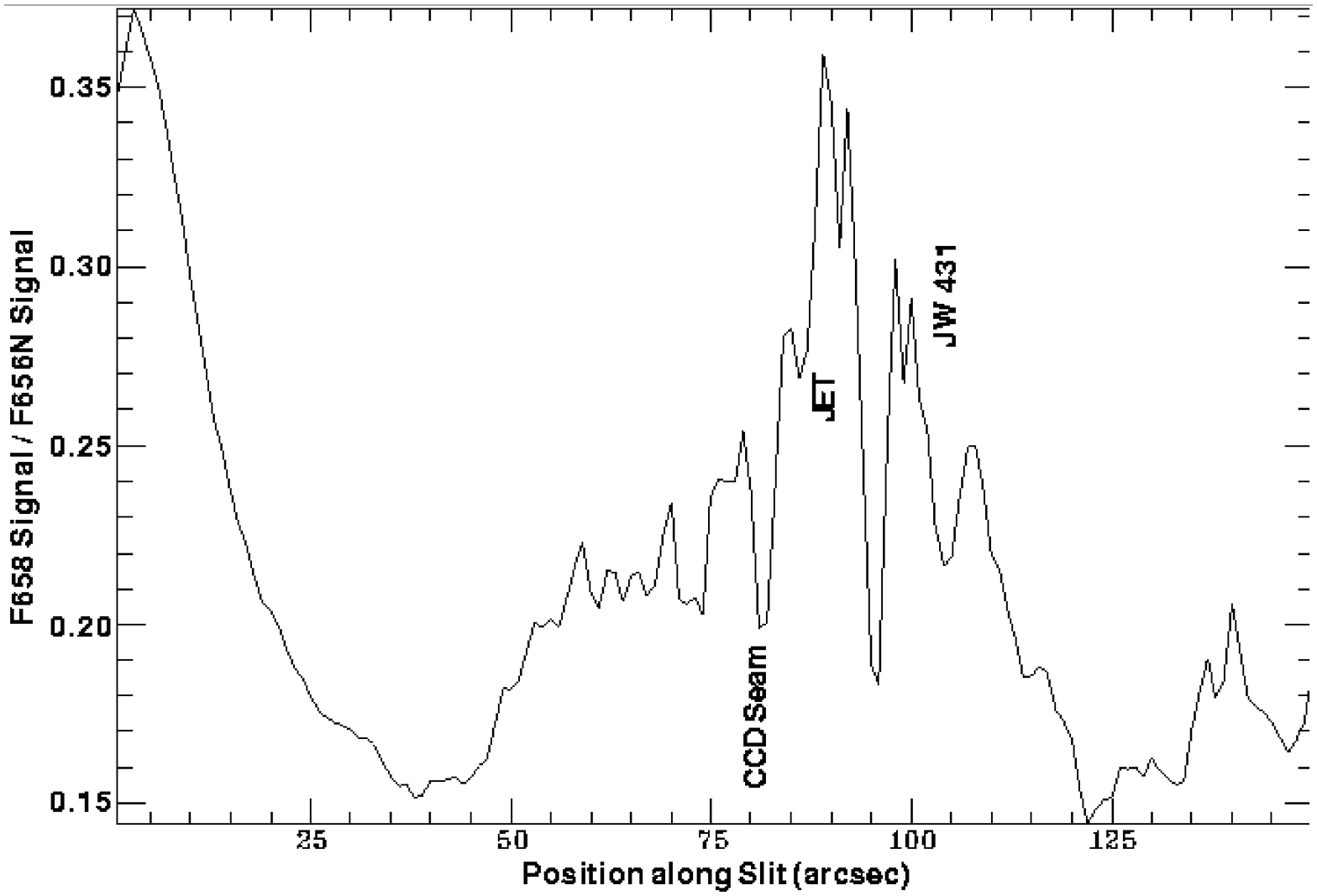]{This is a profile from the left hand portion of Figure 10
over the velocity sample section depicted in Figure 1. The location of the CCD
seam between CCD3 and CCD4 is indicated, as is the location of the only star in
the sample. Both of these features cause uncertainty in the line ratios and
quantitities derived from them.}

\figcaption[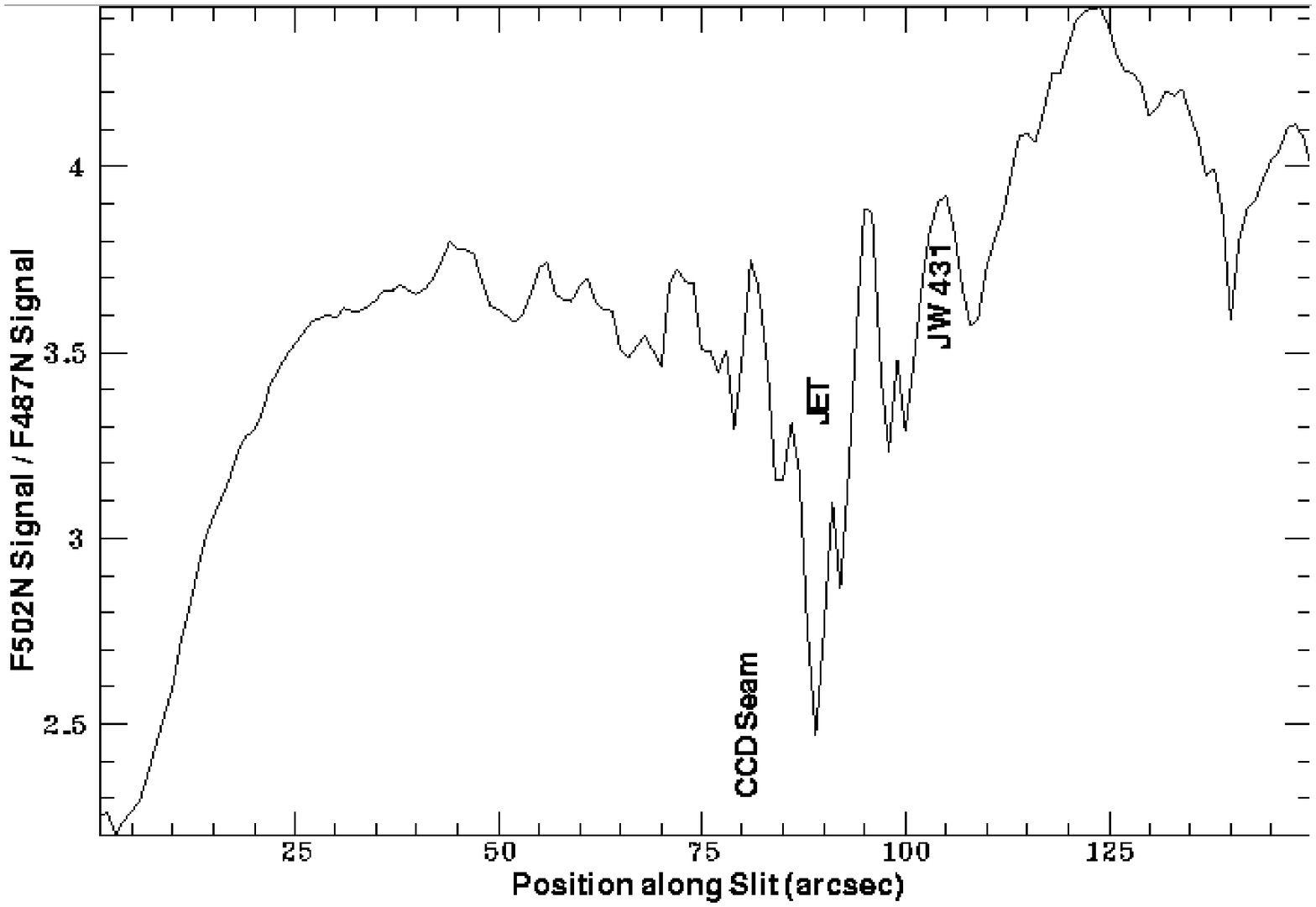]{The same as Figure 11 except for the F658N/F656N ratio of
signals.}

\figcaption[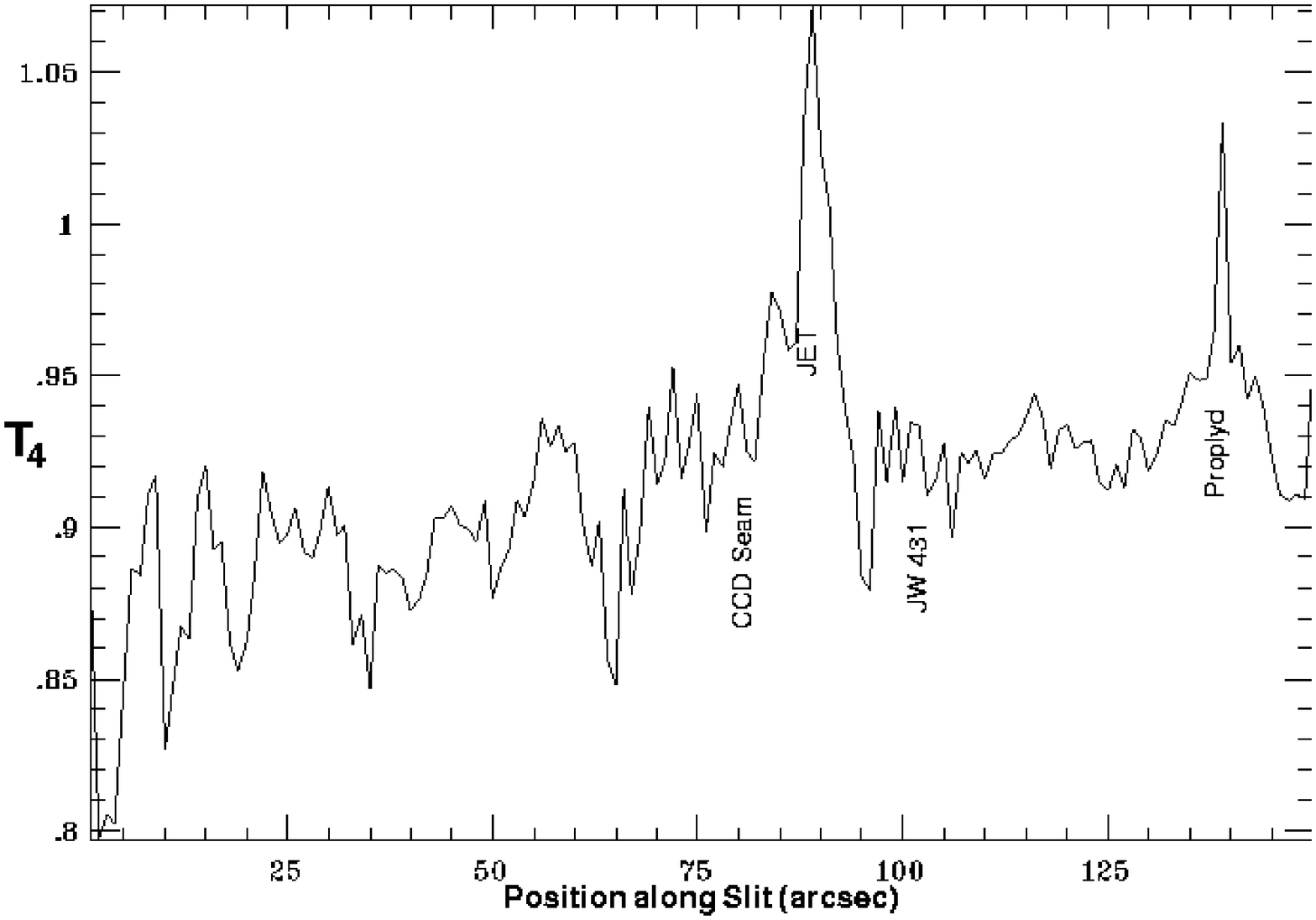]{The electron temperature in units of 10,000 K is shown in
profile for the same section as the velocity samples in Figure 1.}

\figcaption[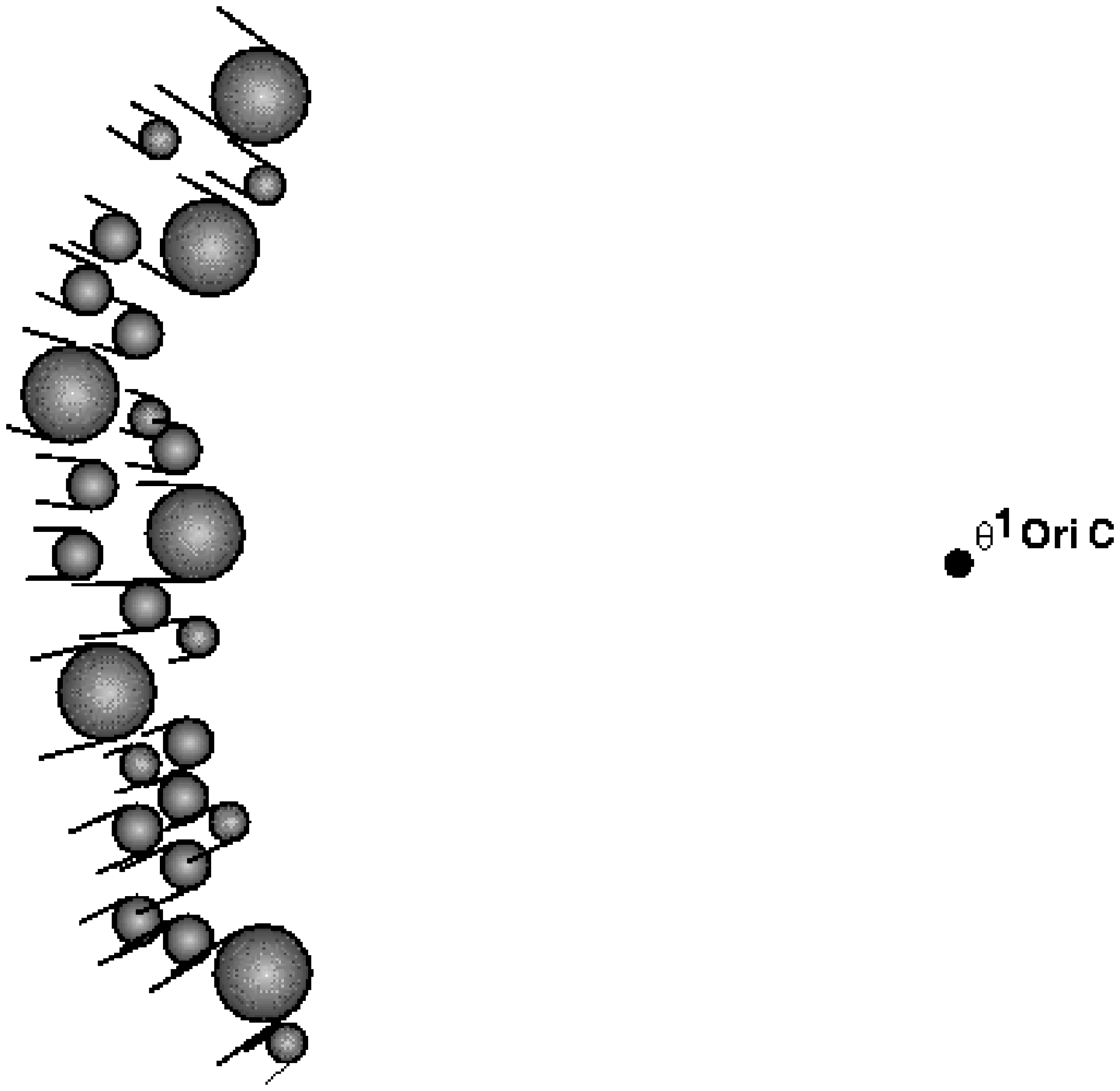]{The proposed knotty model for the ionization front is
presented here in cartoon form.  Each knot would be subject to photoionization
as soon as the average location of the ionization front has moved past it. Each
knot would be brightest on the side facing \thC\ and would be destroyed by the
process of photoevaporation. Prior to their destruction each would have produce
a shadowed region photoionized only by diffuse LyC radiation and having an
electron temperature about three-fifths that of the directly illuminated
material.}

\clearpage

\begin{figure}
\begin{center}
\includegraphics[scale=0.80]{f01.ps}
\end{center}
\end{figure}

\clearpage

\begin{figure}
\begin{center}
\includegraphics[scale=0.80]{f02.ps}
\end{center}
\end{figure}

\clearpage

\begin{figure}
\begin{center}
\includegraphics[scale=1.00]{f03.ps}
\end{center}
\end{figure}

\clearpage

\begin{figure}
\begin{center}
\includegraphics[scale=0.65]{f04.ps}
\end{center}
\end{figure}

\clearpage

\begin{figure}
\begin{center}
\includegraphics[scale=0.65]{f05.ps}
\end{center}
\end{figure}

\clearpage

\begin{figure}
\begin{center}
\includegraphics[scale=0.65]{f06.ps}
\end{center}
\end{figure}

\clearpage

\begin{figure}
\begin{center}
\includegraphics[scale=0.65]{f07.ps}
\end{center}
\end{figure}

\clearpage

\begin{figure}
\begin{center}
\includegraphics[scale=0.95]{f08.ps}
\end{center}
\end{figure}

\clearpage

\begin{figure}
\begin{center}
\includegraphics[scale=0.60]{f09.ps}
\end{center}
\end{figure}

\clearpage

\begin{figure}
\begin{center}
\includegraphics[scale=0.65]{f10.ps}
\end{center}
\end{figure}

\clearpage

\begin{figure}
\begin{center}
\includegraphics[scale=0.65]{f11.ps}
\end{center}
\end{figure}

\clearpage

\begin{figure}
\begin{center}
\includegraphics[scale=0.65]{f12.ps}
\end{center}
\end{figure}

\clearpage

\begin{figure}
\begin{center}
\includegraphics[scale=0.65]{f13.ps}
\end{center}
\end{figure}

\clearpage

\begin{figure}
\begin{center}
\includegraphics[scale=1.00]{f14.ps}
\end{center}
\end{figure}

\clearpage
\end{document}